\documentclass[journal]{IEEEtran}

\usepackage{hyperref}
\usepackage{graphicx} 
\usepackage{booktabs} 
\usepackage{amsmath}  
\usepackage{verbatim}
\usepackage{color}
\definecolor{m_blue}{rgb}{0,0,0}
\usepackage{float}
\usepackage{stfloats} 
\usepackage{multirow} 
\usepackage{hhline} 

\usepackage[space]{cite} 

\begin{document}

\title{Value of Point-of-load Voltage Control for Enhanced \\ Frequency Response in Future GB Power System}

\author{Jinrui Guo,~\IEEEmembership{Student Member,~IEEE,}
        Luis Badesa,~\IEEEmembership{Student Member,~IEEE,}
        Fei Teng,~\IEEEmembership{Member,~IEEE,}
        \\Balarko Chaudhuri,~\IEEEmembership{Senior Member,~IEEE,}
        Shu~Yuen~Ron~Hui,~\IEEEmembership{Fellow,~IEEE}
        Goran Strbac,~\IEEEmembership{Member,~IEEE}
\thanks{This research has been supported by the Engineering and Physical Science Research Council (EPSRC) (Grant EP/K036327/1), IDLES and ABC projects (Grants EP/R045518/1 and EP/S016627/1) and the Hong Kong Research Grant
Council through the Theme-based project under Grant T23-701/14-N. 

J. Guo, L. Badesa, F. Teng, B. Chaudhuri, S.Y.R. Hui and G. Strbac are with the Department
of Electrical and Electronic Engineering, Imperial College London, London,
UK, SW7 2AZ. S.Y.R. Hui is also with the Department of Electrical and Electronic Engineering, The University of Hong Kong.}}

\markboth{IEEE Transactions on Smart Grid, June~2020}%
{Shell \MakeLowercase{\textit{et al.}}: Bare Demo of IEEEtran.cls for IEEE Journals}

\maketitle

\begin{abstract}
The need for Enhanced Frequency Response (EFR) is expected to increase significantly in future low-carbon Great Britain (GB) power system. One way to provide EFR is to use power electronic compensators (PECs) for point-of-load voltage control (PVC) to exploit the voltage dependence of loads. This paper investigates the techno-economic feasibility of such technology in future GB power system by quantifying the total EFR obtainable through deploying PVC in the urban domestic sector, the investment cost of the installment and the economic and environmental benefits of using PVC. The quantification is based on a stochastic domestic demand model and generic medium and low-voltage distribution networks for the urban areas of GB and a stochastic unit commitment (SUC) model with constraints for secure post-fault frequency evolution is used for the value assessment. Two future energy scenarios in the backdrop of 2030 with `smart' and `non-smart' control of electric vehicles and heat pumps, under different levels of penetration of battery energy storage system (BESS) are considered to assess the value of PEC, as well as the associated payback period. It is demonstrated that PVC could effectively complement BESS towards EFR provision in future GB power system.  
\end{abstract}


\IEEEpeerreviewmaketitle



\section*{\textcolor{m_blue}{Nomenclature}}


\subsection*{\textcolor{m_blue}{List of Acronyms}}
\textcolor{m_blue}{\begin{tabular}{@{}ll}
BESS & ~~~~~~~\hspace{-10pt}Battery storage energy system \\
CDC & ~~~~~~~\hspace{-10pt}Cluster of domestic customers\\
EFR & ~~~~~~~\hspace{-10pt}Enhanced Frequency Response\\
EV & ~~~~~~~\hspace{-10pt}Electric vehicle\\
GB & ~~~~~~~\hspace{-10pt}Great Britain\\
GnW & ~~~~~~~\hspace{-10pt}Green World\\
HP & ~~~~~~~\hspace{-10pt}Heat pump\\
LV & ~~~~~~~\hspace{-10pt}Low voltage\\
MV & ~~~~~~~\hspace{-10pt}Medium voltage\\
NSC/SC & ~~~~~~~\hspace{-10pt}Non-Smart/Smart Case\\
PEC & ~~~~~~~\hspace{-10pt}Power electronic compensators\\
PFR & ~~~~~~~\hspace{-10pt}Primary Frequency Response\\
PVC & ~~~~~~~\hspace{-10pt}Point-of-load voltage control\\
SwP & ~~~~~~~\hspace{-10pt}Slow Progression\\
SUC & ~~~~~~~\hspace{-10pt}Stochastic unit commitment\\
VCS & ~~~~~~~\hspace{-10pt}Voltage control at the substation\\
\end{tabular}}

\subsection*{\textcolor{m_blue}{Indices and Sets}}
\textcolor{m_blue}{\begin{tabular}{@{}ll}
$n,~\mathcal{N}$ & ~~~~~~~\hspace{-10pt}Index Set of nodes in the scenario tree \\
$g,~\mathcal{G}$ & ~~~~~~~\hspace{-10pt}Index, Set of generators\\
$s,~\mathcal{S}$ & ~~~~~~~\hspace{-10pt}Index, Set of storage units\\
\end{tabular}}

\subsection*{\textcolor{m_blue}{Constants and Parameters}}
\textcolor{m_blue}{\begin{tabular}{@{}ll}
$\Delta \tau(n)$ & \hspace{-10pt}Time-step corresponding to node $n$ (h)\\
$\pi(n)$ & \hspace{-10pt}Probability of reaching node $n$\\
$c^{\mathrm{ls}}$ & \hspace{-10pt}Load shed penalty (\pounds/MWh)\\
$\mathrm{c}_{g}^{\mathrm{st}}$ & \hspace{-10pt}Startup cost of generating units $g$ (\pounds)\\
$\mathrm{c}_{g}^{\mathrm{nl}}$ & \hspace{-10pt}No-load cost of generating units $g$ (\pounds/h)\\
$\mathrm{c}_{g}^{\mathrm{m}}$ & \hspace{-10pt}Marginal cost of generating units $g$ (\pounds/MWh)\\
$f_0$ & \hspace{-10pt}Nominal frequency of the power grid (Hz)\\
$\rm H_g$ & \hspace{-10pt}Inertia constant of generating units $g$ (s)\\
$\rm H_L$ & \hspace{-10pt}Inertia constant of tripped generator (s)\\
$\rm P_g^{max}$ & \hspace{-10pt}Maximum power output of units $g$ (MW)\\
$\rm P_L^{max}$ & \hspace{-10pt}Largest power infeed (MW)\\
$\operatorname{RoCoF}_{\max}$ & \hspace{-10pt}Maximum admissible RoCoF (Hz/s)\\
$\Delta f_{\mathrm{max}}$ & \hspace{-10pt}Maximum admissible frequency deviation (Hz)\\
$\rm P^D$ & \hspace{-10pt}Total demand besides urban domestic  \\&\hspace{-10pt}customers with PVC (MW)\\
$\rm P^{WN}$ & \hspace{-10pt}Total available wind generation (MW)\\
$\rm P_{min}^{PVC}$ & \hspace{-10pt}Lower bound for $P^{\rm PVC}$ (MW) \\
$\rm P_{max}^{PVC}$ & \hspace{-10pt}Upper bound for $P^{\rm PVC}$ (MW)\\
$\rm T_g$ & \hspace{-10pt}Delivery time of PFR (s)\\
$\rm T_e$ & \hspace{-10pt}Delivery time of EFR (s)\\
\end{tabular}}

\subsection*{\textcolor{m_blue}{Decision Variables}}
\textcolor{m_blue}{\begin{tabular}{@{}ll}
$N_g^{\rm sg}(n)$ &~~~\hspace{-2.5pt}Number of units $g$ that start generating in node $n$\\ 
$N_g^{\rm up}$ &~~~\hspace{-2.5pt}Number of online generating units of type $g$\\
$H$ &~~~\hspace{-2.5pt}Total system inertia ($\rm MW\cdot s$)\\
$P_g$ &~~~\hspace{-2.5pt}Power produced by generating units $g$ (MW)\\
$P^{\rm LS}$ &~~~\hspace{-2.5pt}Load shed (MW)\\
$P_s$ &~~~\hspace{-2.5pt}Power provided by storage units $s$ (MW)\\
$P^{\rm PVC}$ &~~~\hspace{-2.5pt}Power consumed by urban domestic \\&~~~\hspace{-2.5pt}customers with PVC (MW)\\
$P^{\rm WC}$ &~~~\hspace{-2.5pt}Wind curtailment (MW)\\
$R_{\mathcal{G}}$ &~~~\hspace{-2.5pt}Total PFR from all generators (MW)\\
$R_{\mathcal{E}}$ &~~~\hspace{-2.5pt}Total EFR from all storage units and PVC (MW) \\
\end{tabular}}

\subsection*{\textcolor{m_blue}{PVC-related Variables}}
\begin{tabular}{@{}ll}
$I_L$ & ~~CDC current\\
$N_c$, $N_{h}$& ~~Number of CDCs/Number of domestic \\& ~~customers within a CDC\\
$n_p$, $n_q$ & ~~Aggregate active/reactive power-voltage\\
& ~~sensitivity per CDC\\
$P$, $Q$ & ~~Aggregate active/reactive power per CDC\\
$P_{LC}$ & ~~Power loss incurred in the PEC\\
\end{tabular}

\begin{tabular}{@{}ll}
$\Delta P_{LL}$ & ~~Change in network power loss after PVC/VCS\\
$pf^\prime$ & ~~Power factor after PVC\\
$S_{C}$ & ~~Apparent power processed by the PEC\\
$V_B$, $V^{\prime}_B$ & ~~Feeder-side voltage before/after PVC/VCS\\
$V_C$ & ~~Voltage injected by the PEC\\
\end{tabular}

\section{Introduction} \label{Sec:Intro}

The generation mix of Great Britain (GB) has evolved significantly since 2008, with increasing renewable energy displacing traditional gas and coal generation. The ambition is to operate an electricity system with zero carbon output for periods in year 2025 \cite{zerocarbon}. A major challenge here is to maintain system frequency security in the event of contingencies such as a generation loss, as currently renewables do not provide inertia. To deal with the uncertainties of renewables and reduced system inertia, there is an increased need for frequency response services, which can not be solely provided by conventional part-loaded or fast-start thermal plants for economic and carbon emission considerations. 

Various resources for providing ancillary services from the demand-side have been proposed in previous studies \cite{drysdale2015flexible,VincenzoFeiTCL,motalleb2016providing,mu2012primary,teng2016benefits,teng2015benefits}, including thermostatic loads, distributed energy storage systems and electric vehicles (EVs). However, most of them mainly focused on the demonstration of technical feasibility or capability and did not provide much insights on quantifying the benefits of harnessing frequency response services from the demand-side. The availability and benefits from electric space, water heating, cold and wet appliances to participate in GB balancing market have been analyzed in \cite{drysdale2015flexible}. The overall system benefits of multiple services provision from thermostatically-controlled loads \cite{VincenzoFeiTCL}, electrified transportation and heating \cite{teng2016benefits} as well as generic demand-side response \cite{teng2015benefits} have been investigated using an advanced Stochastic Unit Commitment (SUC) model. Nonetheless, Enhanced Frequency Response (EFR), introduced by National Grid GB in 2017 \cite{NationalGridPLossInertia}, has not been considered in the above study. EFR requires the response to be delivered within one second, which could contribute great value in a low inertia system. 

Demand flexibility can be also achieved by exploiting the voltage dependence of loads. For example, peak demand reduction can be realized by decreasing the voltage of distribution feeders through transformer tap changers at the substation \cite{CVR_PNNL_2010,CLASS_TPWRS17_Nando}. Compared with voltage control at the substations (VCS), voltage control at the points of connection of loads allows about double demand reduction on average as demonstrated in \cite{guo2019flexible}. This advantage would be more prominent during high loading conditions. Besides, point-of-load voltage control (PVC) using power electronic compensators (PECs) with a hardware configuration like Electric Springs \cite{yan2016extending} or Unified Power Quality Conditioners \cite{fujita1998unified} allows much shorter response time, making it an ideal option for providing EFR.  

\textcolor{m_blue}{The Smart Transformer (ST) reported in \cite{ST} can provide frequency response by shaping the load consumption accurately through an online load sensitivity identification-based control. Both ST and PVC proposed in this paper utilize the voltage dependency of the loads for shaping their power consumption via PECs. However, there are two basic differences: 1) An ST is located at the supply point of an LV feeder (i.e. at the secondary substation) and hence, its available margin for voltage reduction could be limited compared to PVC under high and/or non-uniform loading of the LV feeders (as illustrated for the VCS approach in Section \ref{Sec:EFR Calculation}) and 2) An ST is a full-rated device (i.e. rated at diversified peak demand of the substation where ST is installed) while the PEC for PVC is a series voltage injection device which is rated at only a fraction (typically 10-15\%) of the diversified peak demand of the cluster of domestic customers (CDCs).}

\textcolor{m_blue}{Also, the quantification of EFR capability from PVC in this paper uses a bottom-up approach based on a stochastic domestic demand model unlike the online measurement-based load sensitivity estimation in \cite{load_control}. Power-voltage sensitivities of each CDC are aggregated from appliance level to single household and further to the CDC level. Note that the introduction of distributed generation would change the net load but the power voltage sensitivity of the load itself would remain the same.}

\textcolor{m_blue}{Reference \cite{Freq_control_microgrid} shows the feasibility of frequency support in an isolated microgrid via a voltage-based frequency controller within the voltage regulator of generators. Although this is a relatively simple solution, the demand reduction capability could be limited under high and/or nonuniform loading similar to the VCS approach discussed later in Section \ref{Sec:EFR Calculation}.}

The effectiveness of using PVC in demand side for rapid frequency response has been demonstrated in \cite{Dip-TSG17-ES-GB-resp} through a case study on the GB power system. However, time variation of different types of loads and their aggregate power-voltage sensitivity was not considered in \cite{Dip-TSG17-ES-GB-resp} to produce a 24-hour EFR profile. Moreover, the study in \cite{Dip-TSG17-ES-GB-resp} did not reflect the actual line voltage drop as no distribution network was considered. Although these problems were mostly addressed in \cite{guo2019flexible}, the value assessment was over-simplified and conservative as it only considered the reserve holding fee and did not include the competing options (e.g. energy storage) under realistic scenarios. A further investigation of the benefits from PVC demand response is yet to be reported, which is the motivation behind the study presented here. \textcolor{m_blue}{The original contributions of this paper are summarized below:}

\begin{enumerate}
    \item \textcolor{m_blue}{The EFR provision through PVC in the urban domestic sector across Great Britain (GB) is quantified through bottom-up aggregation. This method considers a stochastic domestic demand model for typical GB households in conjunction with representative urban medium-voltage (MV) and low-voltage (LV) networks in GB.}
    
    \item \textcolor{m_blue}{A novel frequency-secured stochastic system scheduling framework is extended to incorporate PVC under normal and fully controllable modes. This, for the first time, allows quantification of the value of EFR provision from PVC in terms of reduced system operation cost under future scenarios with high penetration of renewables and energy storage.} 
    
    \item \textcolor{m_blue}{Fundamental insight is provided by analyzing the payback period for the investment in deploying PVC under different future scenarios (Green World and Slow Progression) and operation strategies (‘normal’ and ‘fully controllable’ mode of PVC; ‘smart’ and ‘non-smart’ operation of electric vehicles and domestic heat pumps). Notably, PVC under fully controllable mode results in lower system operation cost and shorter payback period, making PVC resilient to a high penetration of other EFR providers such as battery storage.}
\end{enumerate}


\section{Point-of-Load Voltage Control (PVC)} \label{Sec:PVC}

\begin{figure}[!t]
		\begin{center}		
		\includegraphics[width=1.0\columnwidth]{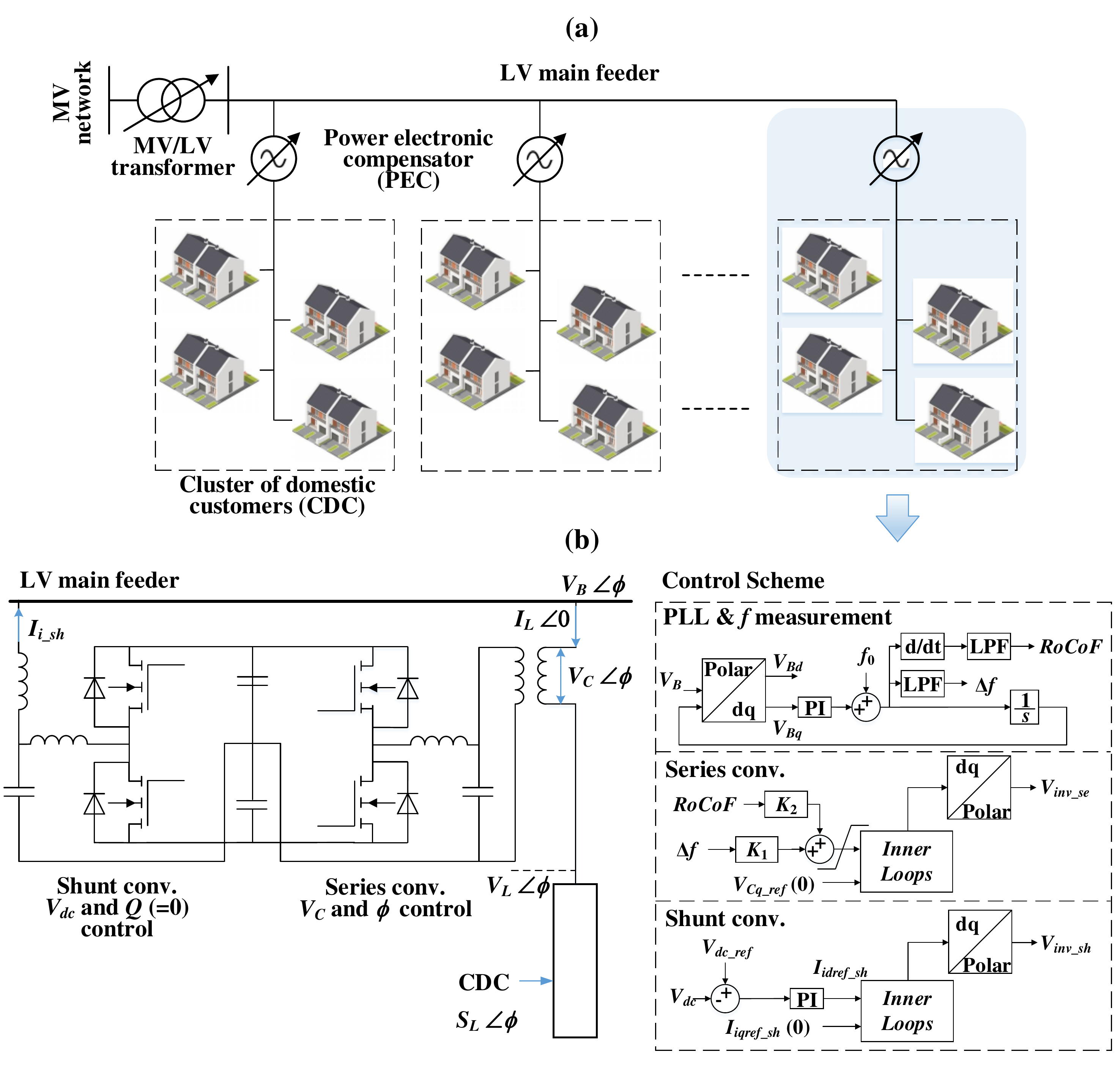}		
		\caption{\textcolor{m_blue}{(a) Concept of point-of-load voltage control (PVC); (b) power electronic compensator (PEC) and control scheme for PVC}}		
		\label{fig:PEC}
		\end{center}
	\end{figure}

PVC can be realized using a PEC which decouples a CDC from an LV feeder, as in Fig.~\ref{fig:PEC} (a). Depending on the configuration of a particular feeder, the number of customers in a CDC could vary. However, the electric proximity within a CDC should be negligible (less than $1\%$) compared to that along the feeder. Thereby, a rural distribution network with long feeders and few customers at each supply point might not be a suitable scenario for PVC application. Besides, due to the lack of an accurate demand model (including demand and power-voltage sensitivity profiles) for industrial and commercial customers, the value assessment of PVC will be limited to urban domestic consumers. Furthermore, PVC may not be necessary to be applied in industrial and commercial sector as the EFR from urban domestic sector is already considerable and the value has a diminishing effect with an increasing level of PVC, as demonstrated later in the paper. 

The voltage of each CDC can be controlled independently within the stipulated range regardless of the feeder-side voltage. This would allow all CDCs to reduce their supply voltages to the minimum acceptable level when demand reduction is required, which offers larger flexibility compared to VCS. The capability of the latter approach would be constrained due to the presence of voltage drop across the feeder.

\textcolor{m_blue}{The PECs used for PVC could use either a single-phase or three-phase version depending on their power rating and location. Either way the PEC is connected in series with the CDC to decouple it from the LV feeder. The detailed circuit diagram of a single-phase PEC \cite{yan2016extending} and its control scheme are shown in Fig.~\ref{fig:PEC} (b). Vector control is used where the feeder-side voltage ($V_B$) is aligned with the d-axis. The magnitude of the injected voltage ($V_C$) is controlled in proportion to the frequency deviation and rate of change of frequency (RoCoF) while the phase angle ($\phi$) is set to be in or out of phase with the feeder-side voltage, both controlled by the series converter. The shunt converter maintains the dc link voltage supporting the active power exchanged by the series converter and operates at unity power factor to minimize the apparent power rating.} Neglecting the converter losses, the active power generated/absorbed by the series converter will be equal to that absorbed/generated by the shunt converter from the feeder. Correspondingly, the total active power consumption will be the same as that of a CDC.  

In the next section, the capability of providing EFR in future GB system will be estimated, as well as the required power rating of PECs considering only urban domestic customers. 

\begin{figure*}[!t]
		\centering	
		\includegraphics[width=2.00\columnwidth]{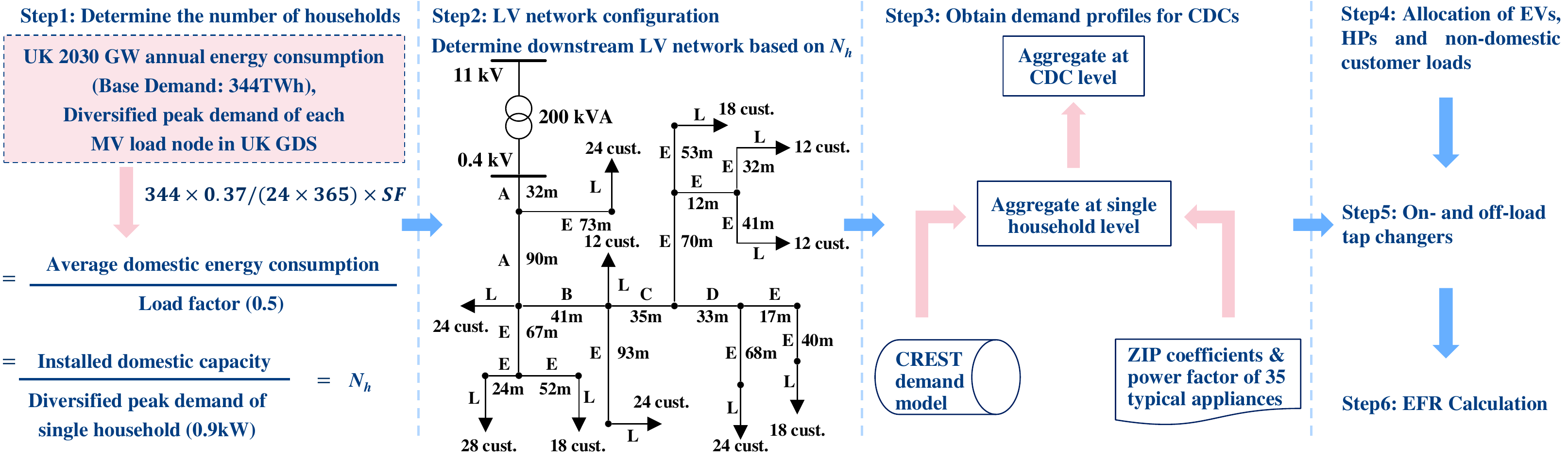}		
		\caption{An outline for estimating EFR from PVC in the urban domestic sector across GB}		
		\label{fig:outline}
\end{figure*}

\section{EFR from PVC in Future GB system} \label{Sec:EFR Calculation}

\subsection{Approach} \label{Sec:approach}

An outline for estimating the EFR provided by PVC in the urban domestic sector in future GB system is given in Fig.~\ref{fig:outline}. The estimation is mainly based on the overall demand profiles, including the base demand along with EV and heat pump (HP) profiles, obtained from the Future Energy Scenarios published by National Grid \cite{NG2030}, high-resolution domestic demand models \cite{CREST-2010} as well as generic MV and LV networks for the GB developed by Center for Sustainable Electricity and Distributed Generation \cite{UKGDS} and CIGRE \cite{Cigre_loadmodel} respectively. It should be noted that no embedded generation is considered in the scenarios for simplicity. A detailed description of the process is as follows:

	\textbf{Step 1:} This part is to determine the number of domestic customers ($N_h$) for each MV node within the GB generic urban distribution network. The average domestic energy consumption for each MV node is obtained by scaling down from the total base demand across GB by a factor (denoted by SF) which is the ratio between the diversified peak demand of individual MV node and that of the total base demand. The installed domestic capacity is then calculated using the average domestic energy consumption (assuming 37\% of the total coming from domestic sector \cite{37}) and a load factor of 0.5 \cite{navarro2012learning}. $N_h$ for a corresponding node is obtained by dividing the average diversified peak demand of a single household (0.9kW) \cite{ELEXON}. 
	
	\textbf{Step 2:} The downstream LV network configuration is then designed based on the calculated $N_h$. To be more specific, the generic `highly urban network' and `urban network' from \cite{Cigre_loadmodel} are adjusted by taking a subset of the whole network to fit in the determined $N_h$. An example of an LV network is as shown in Fig.~\ref{fig:outline}. Each CDC is then formed by 12 to 48 domestic customers \cite{Cigre_loadmodel} at every single LV node.   
	
	\textcolor{m_blue}{\textbf{Step 3:} The overall demand profile for a single household, including the consumption profile of each appliance can be generated by executing the stochastic demand model \cite{CREST-2010, CREST} repeatedly. This model, developed by the Center for Renewable Energy Systems Technology (CREST), generates the daily utilization profile of 35 commonly used appliances in a typical household in GB stochastically. The demand profile is based on a combination of active occupancy pattern and daily activities for a specified number of occupants, day of the week and month of the year. The demand profile and the coefficients of the ZIP load model of each of the 35 appliances \cite{ZIP} are then used to aggregate the demand and power-voltage sensitivity from appliance level to single domestic customer and then to a CDC.}

\begin{figure}[!t]
		\centering	
		\includegraphics[width=0.95\columnwidth]{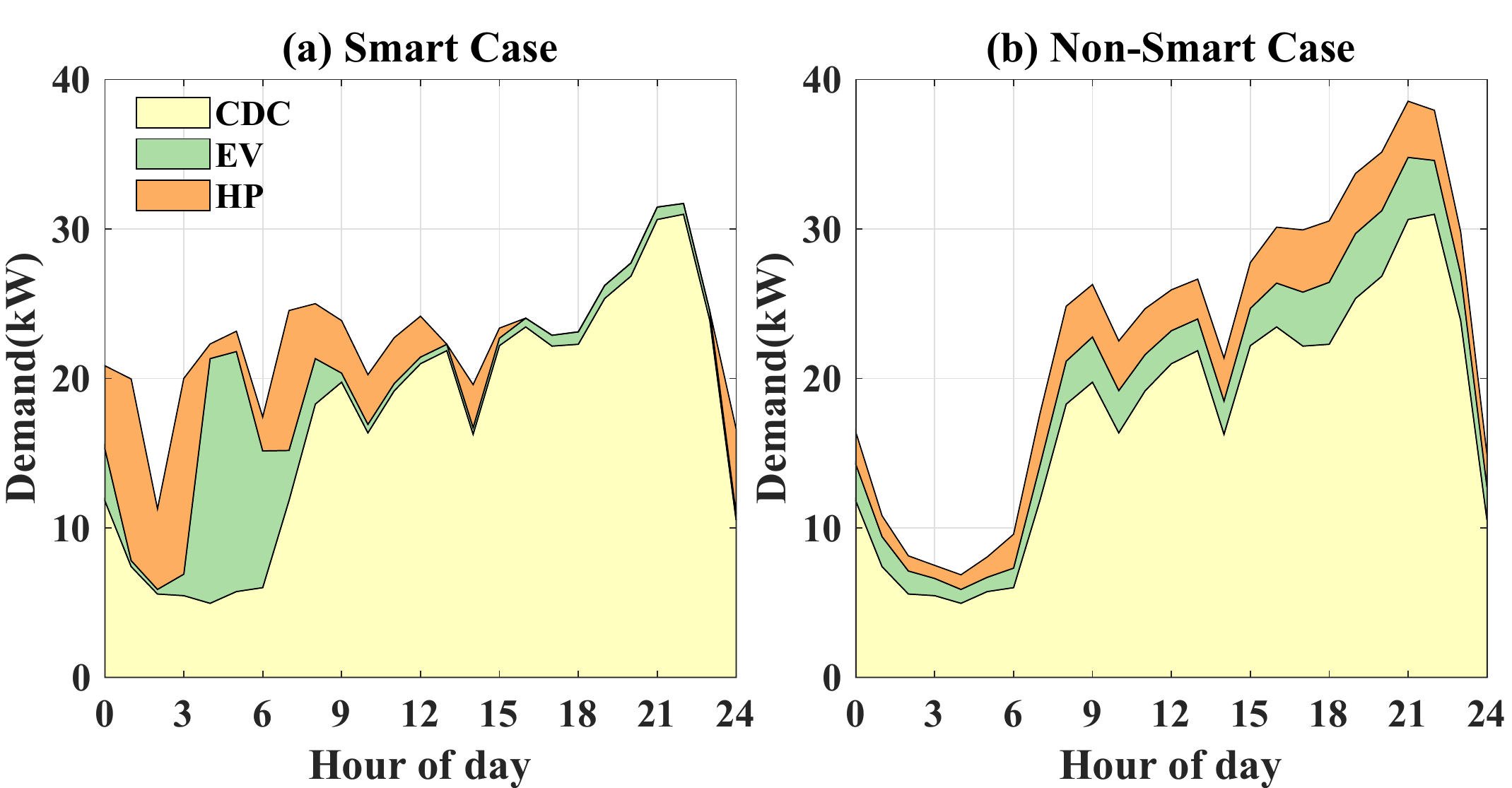}		
		\caption{Demand profiles of a CDC as well as EV and HP loads for a typical summer day (a) Smart Case (SC); (b) Non-Smart Case (NSC)}		
		\label{fig:sanity_check}
	\end{figure}

	\textbf{Step 4:} Taking the domestic demand profiles out of the scaled-down base demand of each MV node, the rest are industrial and commercial customers, all located at the 11 kV side. The EV and HP loads are also scaled down according to the number of domestic customers at each LV node. The original EV and HP profiles are almost in line with the trend of a CDC, referred to as the `Non-Smart Case (NSC)'. Another scenario, called the `Smart Case (SC)' in which the EV and HP consumption are optimized providing energy arbitrage \cite{teng2016benefits}, is also developed to represent a more likely situation for the future as in Fig.~\ref{fig:sanity_check}. All industrial, commercial, EV and HP demand are assumed to be non-responsive to voltage change due to lack of power-voltage sensitivity data. 

	\textbf{Step 5:} An on-load tap change scheme is applied to the primary substation to control the secondary voltage within target voltage with a deadband of 1.5\%. The target voltage is designed based on a load drop compensation scheme \cite{LDC} as in Table~\ref{tab:LDC}. Primary voltages are taken from \cite{ZIP} to avoid a detailed modelling of the upstream ($>$ 33 kV) network. For secondary substations, a seasonal off-load tap change scheme is adopted, in which the tap position is +5\% for winter and +2.5\% for the rest of the year.

	\begin{table}[!hbt]
	\centering
	\caption{Target voltage boost considering load drop compensation, where nominal target voltage is 1.0 p.u.}
	\label{tab:LDC}
	\renewcommand{\arraystretch}{1.2}
\resizebox{0.78\columnwidth}{!}{%
\begin{tabular}{l|llll}
\hline
Loading (kVA) & 0-10 & 10-15 & 15-20 & \textgreater{}20 \\
\hline
Boost (\%)    & 0\%  & 1\%   & 1.5\% & 2\%              \\
\hline       
\end{tabular}
}
\end{table}

	\textbf{Step 6:} Based on all the above assumptions and configurations, the EFR provided by PVC for the urban network can be calculated by (\ref{eq:EFR}) considering the demand reduction capability, change in network power loss due to voltage reduction ($\Delta P_{L L}$) along with the total losses (conduction and switching) incurred in PECs ($\sum_{i=1}^{N_{c}} P_{L Ci}$) and scaled up across GB. \textcolor{m_blue}{The standard power flow model for both LV and MV distribution networks are used to ensure that the voltage and line flow constraints are satisfied during EFR provision and the lower limit for voltage reduction is 0.95 p.u.. Besides, the power loss in PECs has been included in (\ref{eq:EFR}) and would not be considered separately in the cost and payback calculations later in Section \ref{Sec:value}.} A conservative correction factor is applied to the final results to account for urban domestic demand only. This factor is determined by population distribution \cite{population} and extreme cases for energy consumption per head (urban vs. rural) \cite{arbabi2016urban}. \textcolor{m_blue}{Note that the exact EFR available from PVC could be smaller during a contingency such as generation outage when the voltages in electric vicinity of the outage are generally lower than normal. However, such voltage reduction is typically confined within a small region and therefore, will only have marginal impact on the overall EFR available from PVC across GB.} 
	\begin{equation}\label{eq:EFR} 
	\mathrm{EFR}_\mathrm{PVC}= \sum_{i=1}^{N_{c}} P_{i 0}\left[V_{B i}^{n_{pi}}-V_{\min }^{n_{p i}}\right] -\Delta P_{L L}-\sum_{i=1}^{N_{c}} P_{L Ci} 
	\end{equation}

\subsection{EFR Capability with PVC} \label{Sec:EFR results}

In this section, the EFR capability of PVC in the urban domestic sector for the SC of 2030 Green World (GnW) scenario is presented and analyzed. The characteristics of this 2030 GnW scenario will be discussed in Section \ref{Sec:simulated scenarios}.

\begin{figure*}[ht]
		\centering	
		\includegraphics[width=1.9\columnwidth]{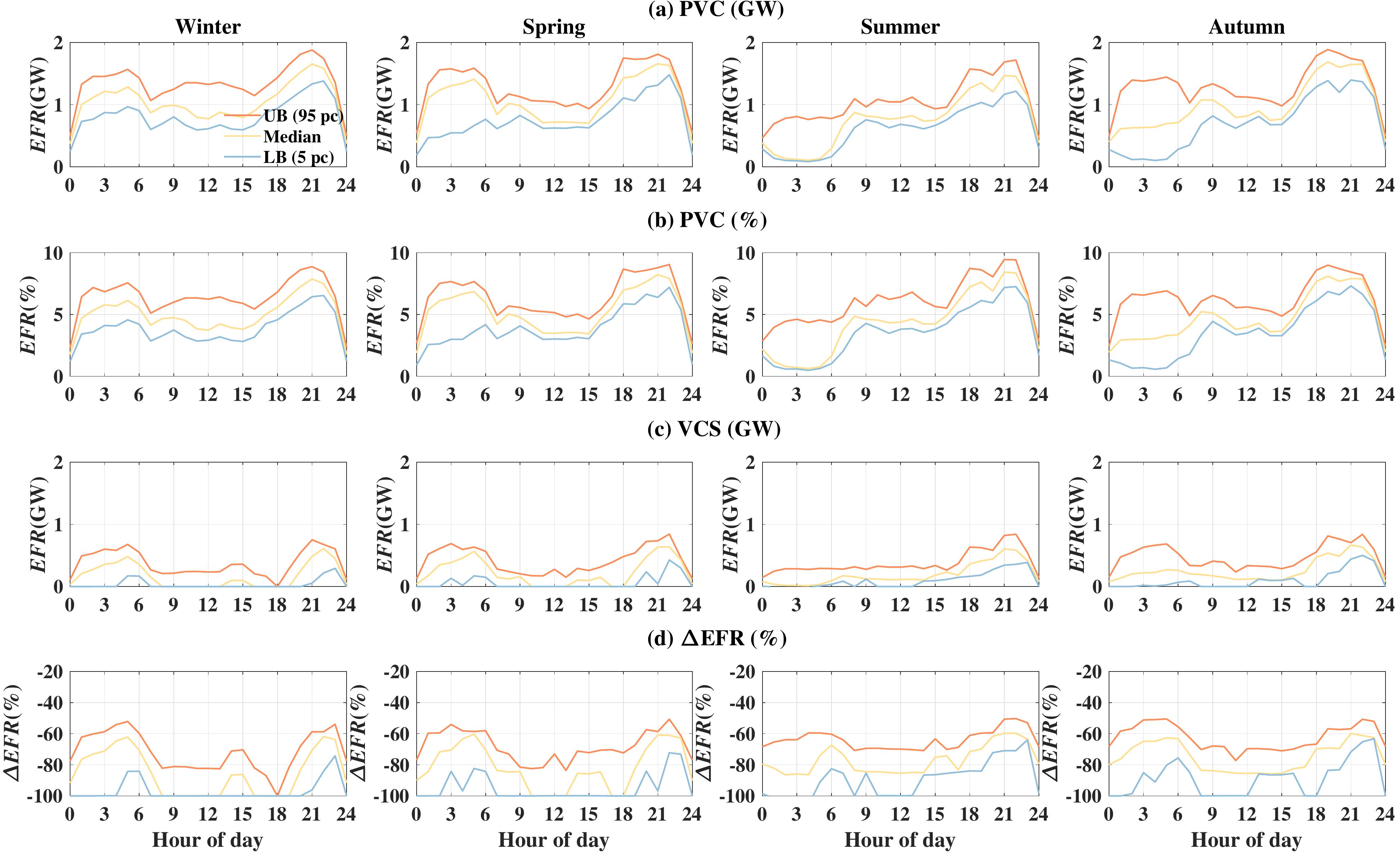}		
		\caption{Comparison between EFR with PVC and EFR with VCS for the 2030 GnW SC}		
		\label{fig:EFR}
	\end{figure*}

The comparison between EFR provision with PVC and that with VCS for the 2030 GnW SC is as shown in Fig.~\ref{fig:EFR}. \textcolor{m_blue}{The EFR achieved from VCS is given by (\ref{eq:VCS}), where $V_{B i}^{\prime n_{p i}}$ corresponds to feeder-side voltage after VCS and is decided by moving the tap changer by 0.005p.u. per step until the minimum feeder-side voltage in the system violates the criteria (0.95p.u.).} The three traces in each subplot represent the upper boundary (UB, 95 pctl.), median value and lower boundary (LB, 5 pctl.) of the results considering daily variations within that season. 
\textcolor{m_blue}{
\begin{equation} \label{eq:VCS}
\mathrm{EFR}_\mathrm{VCS}=\sum_{i=1}^{N_{c}} P_{i 0}\left[V_{B i}^{n_{p i}}-V_{B i}^{\prime n_{p i}}\right]-\Delta P_{LL}
\end{equation}}

EFR with PVC ranges from about 0.1 to 1.9GW throughout the year, accounting for 0.5\% to 9.4\% of the domestic diversified peak demand of the corresponding day. The value peaks during 18:00 to 22:00 for all seasons mainly due to high loading levels along with voltage-reduction room for a considerable number of CDCs. In winter and some days of spring and autumn, there is a significant amount of EFR available between midnight and early morning (6 am) due to the presence of electric space heating and also a relatively large room for voltage reduction. However, the number of households which has got the space heating turned on during this time varies a lot between the days of a particular season which causes the variability (the difference between UB and LB) in EFR even during 12 am $-$ 6 am. This variability would be less if the results are shown for each month.  

Comparatively, the availability of EFR through VCS varies from 0 to about 0.85GW. An index `$\Delta \mathrm{EFR}$' as defined in (\ref{eq:VR}) further quantifies this comparison. As shown in Fig.~\ref{fig:EFR}, the EFR availability from VCS is at least 50\% lower than that from PVC throughout the year. In some instances, $\Delta \mathrm{EFR}$ reaches -100\%, indicating that VCS completely loses its capability of providing EFR as the voltage at the far end of the feeder is below the allowable limit $V_{\rm min}$. 
\begin{equation}\label{eq:VR}
\Delta \mathrm{EFR} (\%) = \frac{\mathrm{EFR}_\mathrm{VCS} - \mathrm{EFR}_\mathrm{PVC}} {\mathrm{EFR}_\mathrm{PVC}}\times100\%
\end{equation}

Besides the lower EFR capability aspect, another two disadvantages of implementing EFR via VCS are: 1) VCS is incapable of handling the case when one of the branches has reverse power flow due to high penetration of distributed photovoltaic; 2) the delivery time of EFR should be within one second, which means VCS would require a sizeable PEC. This would bring many challenges in protection scheme design. 
	
\subsection{Investment in PECs} \label{Sec:Rating}

The rating of the PEC required at the $i\rm{th}$ CDC is determined by rounding up the maximum power processed by the PEC (as given below) to a whole number: 
\begin{equation} \label{eq:PEC}
 \begin{aligned} S_{C i} &=V_{C i} I_{L i}\left(1+p f_{i}^{\prime}\right) \\ &=\left(V_{B i}^{\prime}-V_{\min }\right) \sqrt{P_{i 0}^{2} V_{\min }^{2\left(n_{p i}-1\right)}+Q_{i 0}^{2} V_{\min }^{2\left(n_{q i}-1\right)}}\left(1+p f_{i}^{\prime}\right) \end{aligned}   
\end{equation}
where, $V_{Bi}^\prime$, $pf_i^\prime$ denote the modified feeder-side voltage and power factor, respectively as a result of reducing the voltage of the $i\rm{th}$ CDC to $V_{\rm min}$. 

For the 2030 GnW SC, the required PEC rating ranges from 4 to 14 kVA for each CDC, with a total of about 6.6 GVA across the GB if all urban domestic customers are equipped with PECs for PVC. \textcolor{m_blue}{According to \cite{Converter_Cost}, the typical price of single-phase converters is \$140/kW but could vary from \$60/kW to \$220/kW while the maintenance cost is about \$10/kW per year. The variability in the converter cost (including maintenance cost) is considered to determine the range of payback period in Section \ref{sec:payback} with an exchange rate of \pounds 1=\$1.25. Note that the customers under PVC do not have to be compensated as the supply voltages will be maintained within the stipulated limits. In fact, the PECs used for PVC can help optimize the supply voltage to maximize customer benefit (e.g. PowerPerfector deployed in the commercial sector) when system services (e.g. EFR) are not necessary.}

\section{Value of EFR from PVC in Future GB System}  \label{Sec:value}

In this section, the benefits of PVC in providing EFR under both normal mode and fully controllable mode are quantified, considering daily and seasonal capability variations obtained from Section \ref{Sec:EFR results}. The fully controllable mode is referred to the case in which the demand can be controlled to consume more/less as necessary to provide more/less EFR. 

\subsection{Methodology} \label{Sec:SUC}

The frequency-constrained SUC model proposed in \cite{LuisEFR} is extended and applied here for assessing the value of implementing PVC in the urban domestic sector. The model simultaneously optimizes energy production as well as provision of primary frequency response (PFR) and EFR in the light of wind generation uncertainties.

\begin{figure}[t]
		\centering	
		\includegraphics[width=0.8\columnwidth]{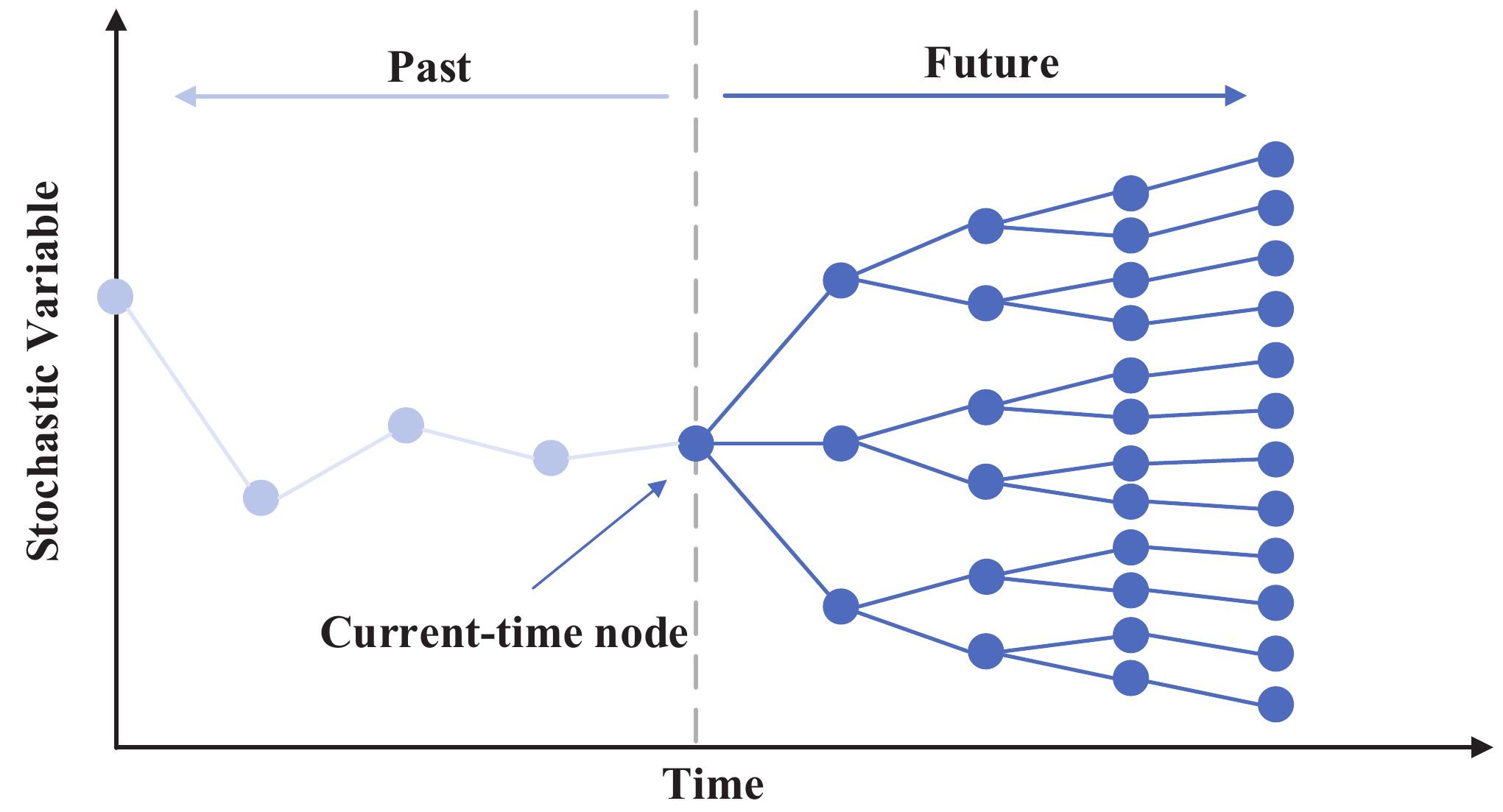}		
		\caption{Example of a scenario tree used in the SUC model \cite{AlexEfficient}}		
		\label{fig:tree}
	\end{figure}

Uncertainties are modelled using a scenario tree, which represents user-defined quantiles of the distribution of the stochastic variable as described in \cite{AlexEfficient}. A rolling scheduling is applied here: for each time step ($\Delta \tau(n)=1\rm h$ in our case), a scenario tree covering a 24-hour horizon is built and the SUC optimization over all nodes in the scenario tree is computed correspondingly. Only the here-and-now decisions in the current-time node are preserved with all future decisions discarded. This process is repeated for each time step. A schematic diagram of a scenario tree is shown in Fig.~\ref{fig:tree}.

The objective is to minimise the expected operation cost:
\begin{equation}\label{eq:op}  
\sum_{n \in \mathcal N} \pi(n)\left(\sum_{g \in \mathcal G} C_{g}(n)+\Delta \tau(n)\left[c^{\mathrm{ls}} P^{\mathrm{LS}}(n)\right]\right)    
\end{equation}
in which the operating cost of generator $g$ is given by (\ref{eq:op}) and the latter term represents the load shed penalty. Note that the variables in (\ref{eq:op}) to (\ref{eq:F21}) in italics represent decision variables while the others denote constants and parameters. $\pi(n)$ and $\Delta \tau(n)$ are the probability of reaching node $n$ and the time-step corresponding to node $n$. $c^{\mathrm{ls}}$, $c_g^{\mathrm{st}}$, $c_g^{\mathrm{nl}}$ and $c_g^{\mathrm{m}}$ stand for load shed penalty, startup, no-load and marginal cost of generating units $g$ while $P^{\mathrm{LS}}$, $P_g$, $N_g^{\mathrm{sg}}$ and $N_g^{\mathrm{up}}$ denote load shed, power produced by unit $g$ and the number of units start generating and online in node $n$ respectively. 
\begin{equation}
C_{g}(n)=\mathrm{c}_{g}^{\mathrm{st}} N_{g}^{\mathrm{sg}}(n)+\Delta \tau(n)\left[\mathrm{c}_{g}^{\mathrm{nl}} N_{g}^{\mathrm{up}}(n)+\mathrm{c}_{g}^{\mathrm{m}} P_{g}(n)\right]
\end{equation}

The objective function is subject to a series of constraints. Only the load-balance and frequency-security constraints are listed here as they are directly associated with the introduction of PVC. For all other operating constraints of thermal plants and storage, readers are advised to refer to the Appendix of \cite{FeiStochastic}. Index $n$ is omitted in following equations for simplicity. 

The load-balance constraint is given by:
\begin{equation} 
\sum_{g \in \mathcal G} P_{g}+\sum_{s \in \mathcal S} P_{s} +{\rm P}^{\mathrm{WN}}-P^{\mathrm{WC}} = {\rm P}^{\rm D}+P^{\rm PVC}-{ P}^{\mathrm{LS}} 
\end{equation}
where $\rm P^D$ denotes all demand in the system except the urban domestic customers with PVC; $\rm P^{WN}$, $P^{\rm WC}$ and $P_s$ are total available wind power, wind curtailment and power provided by storage unit $s$. In the normal mode, $\rm P^{PVC}$ is the demand equipped with PVC and it is a constant for each node while in the fully controllable mode, the demand becomes a decision variable ($P^{\rm PVC}$) within boundaries of $[\rm P^{PVC}_{min},P^{PVC}_{max}]$. 

Frequency security constraints that guarantee RoCoF, quasi-steady-state and nadir requirements are given by equations (\ref{eq:F1}) to (\ref{eq:F3}), in which $\rm P_L^{max}$ and $\rm H_L$ are the largest power infeed and the corresponding inertia constant, as deduced in \cite{LuisEFR}:
\begin{equation} \label{eq:F1}
H =\sum_{g \in \mathcal{G}} {\rm H}_{g} \cdot {\rm P}_{g}^{\max} \cdot N_{g}^{\mathrm{up}}- {\rm P_{L}^{\max}} \cdot {\rm H_{L}} \geq\left|\frac{{\rm P_{L}^{\max}} \cdot f_{0} }{2 \operatorname{RoCoF}_{\max}}\right|
\end{equation}
\begin{equation} \label{eq:F2}
R_{\mathcal{G}}+R_{\mathcal{E}} \geq {\rm P_{L}^{\max}}
\end{equation}

\begin{equation} \label{eq:F3}
\left(\frac{H}{f_{0}}-\frac{R_{\mathcal{E}} \cdot \mathrm{T}_{\mathrm{e}}}{4 \cdot \Delta f_{\mathrm{max}}}\right) \cdot R_{\mathcal{G}} 
\geq \frac{\left({\rm P_{L}^{\max}}-R_{\mathcal E}\right)^{2} \cdot \mathrm{T}_{\mathrm{g}}}{4 \cdot \Delta f_{\max }} 
\end{equation}
where:
\begin{equation} \label{eq:F21}
R_{\mathcal{E}} = {\rm EFR_{PVC}} + {\rm EFR_{S}}
\end{equation}
\textcolor{m_blue}{\begin{equation} \label{eq:F22}
\mathrm{EFR}_{\mathrm{S}} \leq \bar{P}_{\mathrm{s}}-P_{\mathrm{s}}
\end{equation}}

$R_{\mathcal G}$ corresponds to total PFR from all generators while EFR ($R_{\mathcal E}$) is provided by both storage units and urban domestic demand with PVC, as in (\ref{eq:F21}), where the maximum value of $\rm EFR_{PVC}$ is obtained by (\ref{eq:EFR}) as described in Section \ref{Sec:EFR Calculation} \textcolor{m_blue}{and $\rm EFR_{S}$ is limited by the difference between maximum discharge rate ($\bar{P}_{\mathrm{s}}$) and actual discharge rate/power output (${P}_{\mathrm{s}}$). For example, a BESS with 0.5GW capacity can provide up to 1GW of EFR if it is fully charging (i.e. ${P}_{\mathrm{s}} = -0.5\mathrm{GW}$), and 0GW if it is fully discharging.} Due to the non-linearity of (\ref{eq:F3}), a linearization method is required to be implemented in a Mixed-Integer Linear Program. Readers are advised to refer to \cite{LuisEFR} for further details. 

With the above methodology, the value and the corresponding payback of PVC is quantified by computing the reduction in system operating cost. The benefits in $\rm CO_2$ emission and wind curtailment reduction are also presented.

\subsection{Scenarios} \label{Sec:simulated scenarios}

Two future scenarios for GB's generation and demand, i.e. 2030 Slow Progression (SwP) and 2030 Green World (GnW), are considered here. The installed wind capacity is of 72.8 and 41.8GW for 2030 GnW and SwP, respectively \cite{teng2016benefits}. The configuration and characteristics of the main thermal plants are summarized in Table \ref{tab:char}. A biomass plant of 1.75GW-rating is also present in both scenarios, while five additional coal units with Carbon Capture and Storage capabilities are also included in the 2030 GnW case, with a rating of 0.7GW each \cite{teng2016benefits}. 

\begin{table}[!h]

\caption{Characteristic of Main Thermal Plants \cite{teng2016benefits}} 
\label{tab:char} 
\renewcommand{\arraystretch}{1.2}
\resizebox{\columnwidth}{!}{%
\begin{tabular}{l|cccc}
\hline
\hline
\multicolumn{1}{c|}{}       & Nuclear & CCGT  & OCGT   & Coal  \\
\hline
Number of Units (2030 SwP)  & 5       & 94    & 33     & 0     \\
Number of Units (2030 GnW)  & 6       & 70    & 75     & 4     \\
Rated Power (MW)           & 1800    & 467   & 205    & 836   \\
Min Stable Generation (MW) & 1800    & 233.5 & 82     & 292   \\
No-Load Cost (\pounds/h)         & 391     & 2641  & 11328  & 4474  \\
Marginal Cost (\pounds/MWh)      & 4.82    & 68.75 & 195.12 & 86.6  \\
Startup Cost (\pounds)           & 49362.5 & 32000 & 0      & 21000 \\
Startup Time (h)           & N/A     & 4     & 0      & 4     \\
Min Up Time (h)            & N/A     & 4     & 1      & 4     \\
Min Down Time (h)          & N/A     & 0      & 1      & 0       \\
Inertia Constant (s)       & 5       & 5     & 5      & 5     \\
Max PFR deliverable (MW)   & 0       & 233.5   & 40     & 300   \\
Emissions (kgCO2/MWh)      & 0       & 394   & 557    & 925  \\
\hline
\hline
\end{tabular}
 
}

\end{table}

Besides, a pumped storage unit with 10GWh capacity, 2.6GW rating and 75\% efficiency is also present in both scenarios, corresponding to the Dinorwig unit in GB. Along with PVC in the urban domestic sector, Battery Energy Storage Systems (BESS) with a 2.5GWh tank, 0.5GW rating and 90\% round-trip efficiency also serves to provide EFR. The delivery time of EFR ($\rm T_e$) and PFR ($\rm T_g$) are 0.5s and 10s respectively. The largest infeed power loss ($\rm P^{max}_L$) is 1.8GW and the load shed penalty ($c^{\mathrm{ls}}$) is \pounds 50k/MWh. Dynamic frequency requirements are:  $\Delta f_\mathrm{max}$ = 0.8Hz, $\Delta f^\mathrm{ss}_\mathrm{max}$ = 0.5Hz and $\rm RoCoF_{max}$ = 0.5Hz/s. The optimization problem was solved with FICO Xpress 8.0, linked to a C++ application via the BCL interface.
 
\begin{figure}[!t]
		\centering	
		\includegraphics[width=0.95\columnwidth]{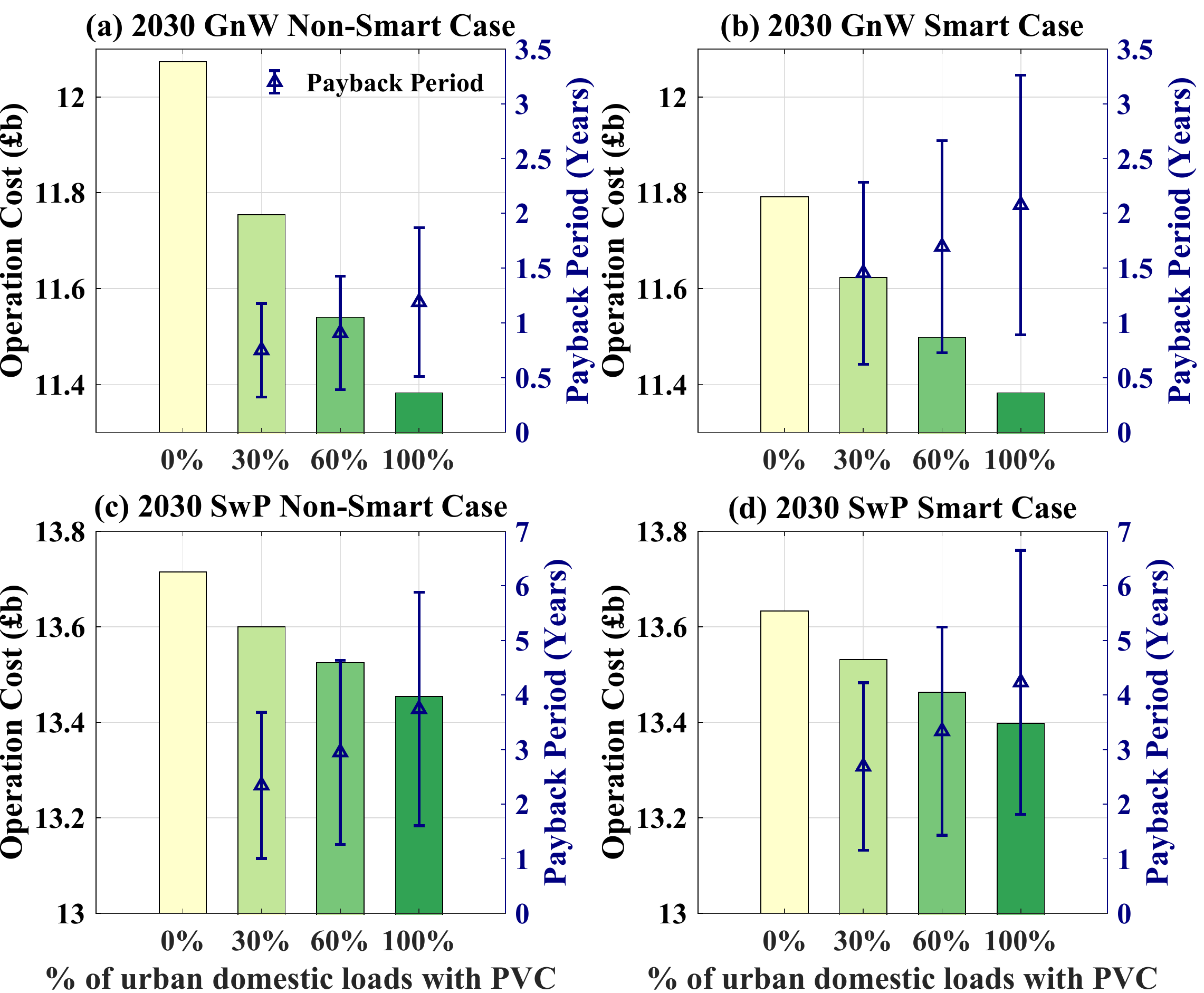}		
		\caption{\textcolor{m_blue}{Operation cost and payback period for (a) 2030 GnW NSC; (b) 2030 GnW SC; (c) 2030 SwP NSC; (d) 2030 SwP SC}}		
		\label{fig:value_nominal}
	\end{figure}

\subsection{PVC in Normal Mode} \label{Sec:Nominal Case}
\textcolor{m_blue}{Under the normal mode, PVC would not be utilized (or activated) i.e. voltage (and hence, actual power consumption) of the CDCs would not actually be reduced to the minimum stipulated limit.} However, the maximum capability of providing EFR is still determined by reducing the voltages at all the CDCs to the minimum stipulated value, as obtained in Section \ref{Sec:EFR results}. In this subsection, the benefits of PVC are shown for different percentages of urban domestic loads under PVC for all scenarios. 

\subsubsection{System Operation Cost} \label{sec:payback}

The operation costs (in \pounds b) for the Non-Smart and Smart cases in the 2030 GnW and SwP scenarios with 0\%, 30\%, 60\% and 100\% of urban domestic customers under PVC are presented in Fig.~\ref{fig:value_nominal}. With 100\% of urban domestic loads with PVC, the maximum operation cost saving can be as high as \pounds 0.72b in the 2030 GnW NSC while the minimum number is about \pounds 0.23b in the 2030 SwP SC. The cost reduction is due to PVC reducing the burden on conventional thermal generators for providing frequency response. The higher economic value observed in the GnW scenario and NSC can be explained as, in these circumstances, there would be larger variations in demand and wind generation, which increases the need for EFR.

Along with the operation cost, the payback period is also given considering the investment in PVC as calculated in Section \ref{Sec:Rating}. \textcolor{m_blue}{Depending on the percentage of loads with PVC, the payback period ranges from 0.3 to 3.3 years for 2030 GnW and 1 to 6.7 years for 2030 SwP, respectively considering a range of converter price.} The increase in payback period with increasing percentage of PVC-based loads clearly implies a saturation effect, i.e. the first megawatts of EFR from PVC generate the highest economic value for the system. \textcolor{m_blue}{It is to be noted that the calculated payback periods are still conservative as the cost savings do not include carbon price, for example. Moreover, only single-phase PECs are considered which have higher price in \$/kW compared to three-phase PECs (as outlined in \cite{Converter_Cost}) which makes the payback period even more pessimistic.}

\begin{figure}[!t]
		\centering	
		\includegraphics[width=0.95\columnwidth]{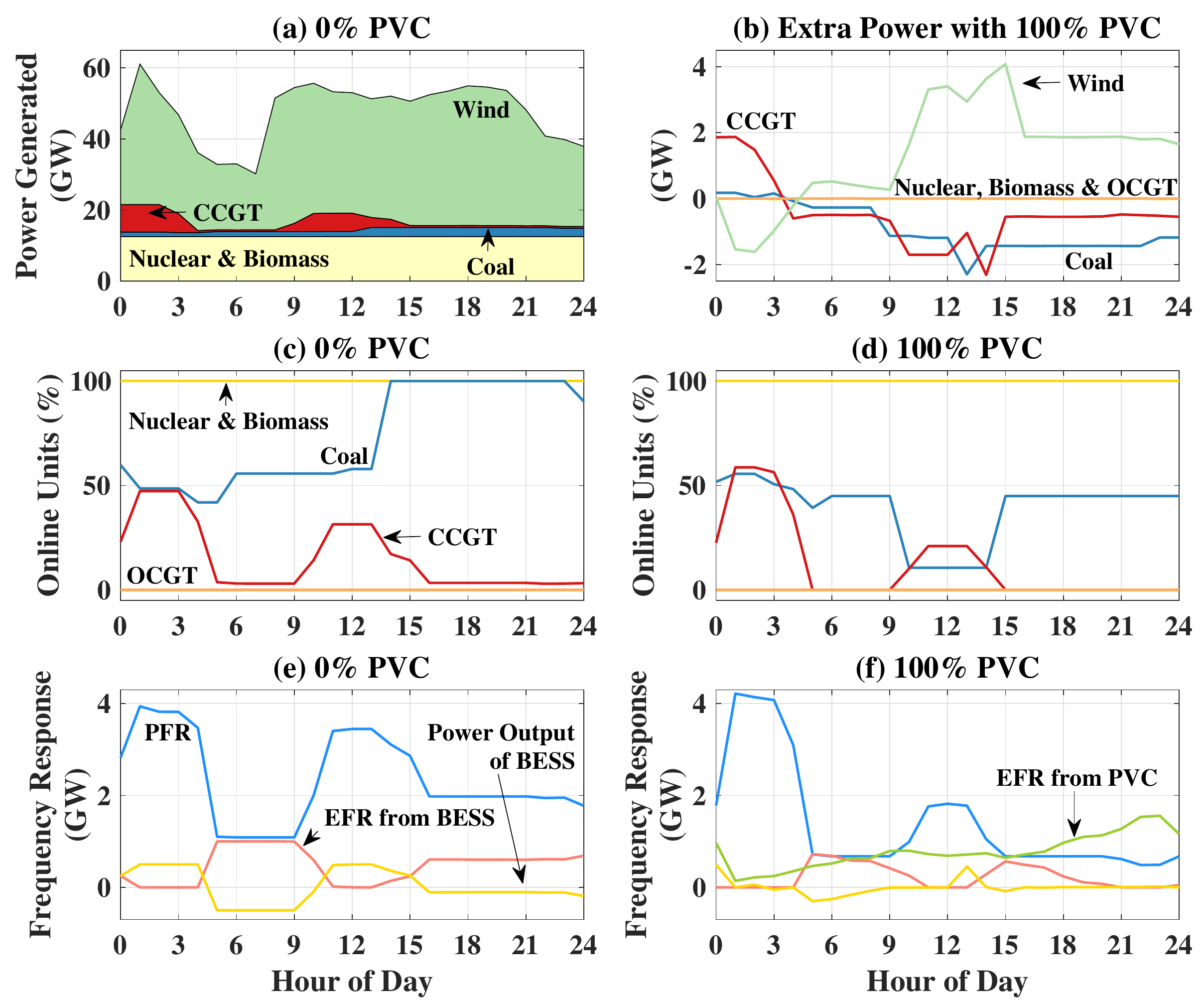}		
		\caption{\textcolor{m_blue}{24-hour dispatch profile for different types of generation and frequency response for a day with low net-demand in 2030 GnW SC}}		
		\label{fig:LND}
	\end{figure}
	
	\begin{figure}[!t]
		\centering	
		\includegraphics[width=0.95\columnwidth]{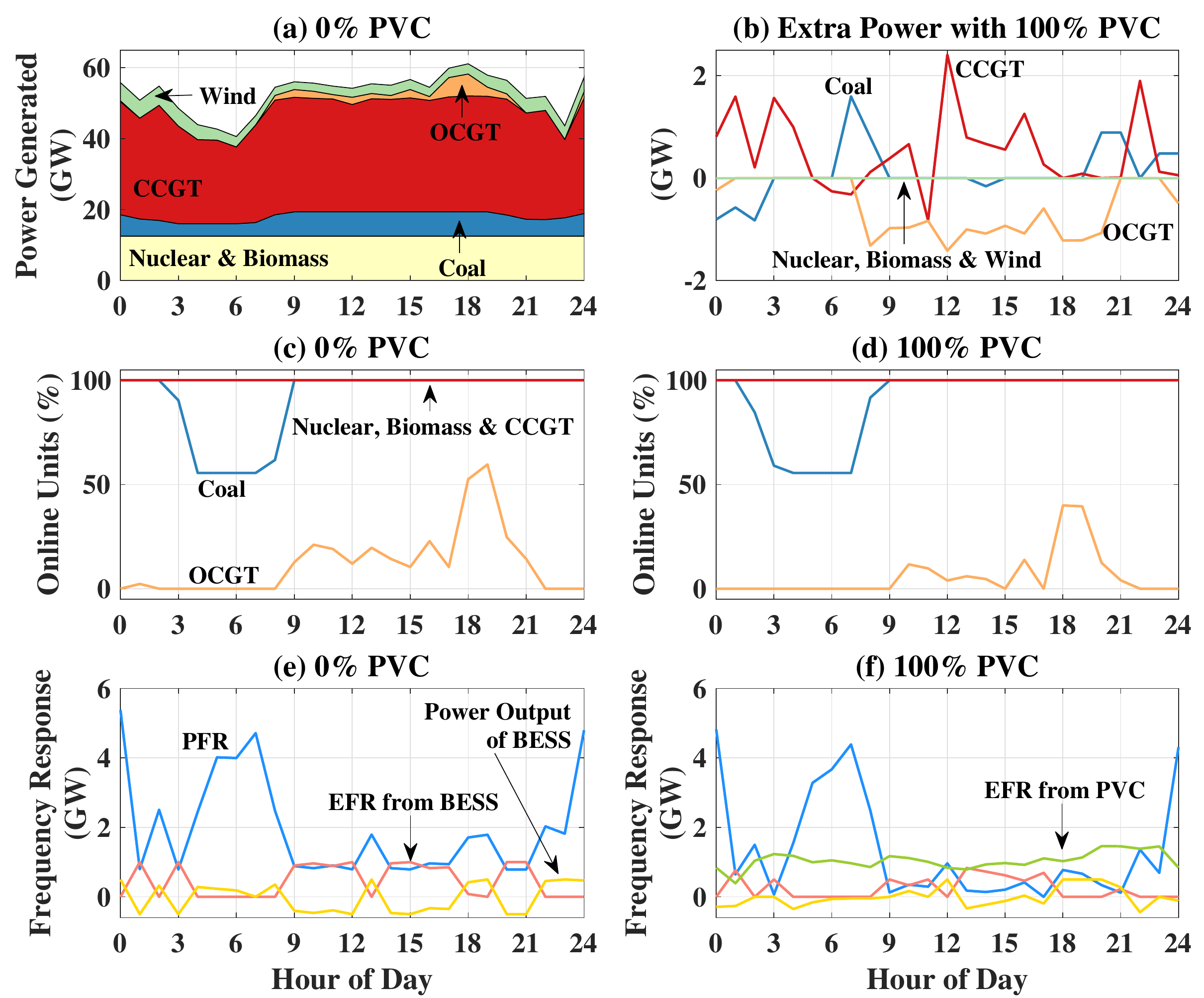}		
		\caption{\textcolor{m_blue}{24-hour dispatch profile for different types of generation and frequency response for a day with high net-demand in 2030 GnW SC}}		
		\label{fig:HND}
	\end{figure}

\textcolor{m_blue}{To further get useful insight into the cost savings realized through PVC, the dispatch profiles for two days with low and high net-demand (demand minus wind accommodated) are shown in Figs. \ref{fig:LND} and \ref{fig:HND}. It is to be noted that subplots \ref{fig:LND} (b) and \ref{fig:HND} (b) represent the extra power generated by each type of generation with 100\% PVC compared to the case without PVC. Following observations can be made from Figs. \ref{fig:LND} and \ref{fig:HND}:
a)	During low net-demand periods in Fig. \ref{fig:LND}, the benefit of using PVC for EFR provision is very high as it reduces the number of online part-loaded coal and CCGT units that are otherwise required to provide PFR to accommodate more wind generation. As shown in Fig. \ref{fig:LND}, both number of online coal and CCGT units and their power output are significantly reduced while more wind is accommodated. 
b)	During high net-demand periods in Fig. \ref{fig:HND}, the reduction in operation cost from implementing PVC is moderate but still noticeable. The cost saving in this case mainly comes from relieving CCGTs from the PFR duty so that less generation from expensive OCGTs is required to supply the load. As shown in Fig. \ref{fig:HND}, the number of online OCGT units and their power output is much less between 8:00 and 22:00 when PVC provides EFR.
c)	The EFR scheduled from BESS is lower in presence of PVC for both high and low net-demand, referred to Figs. \ref{fig:LND} (f) and \ref{fig:HND} (f). This means that the BESS can be used more frequently for energy arbitrage instead of providing EFR, which is another key advantage of the PVC.}

\begin{figure}[!t]
		\centering	
		\includegraphics[width=0.95\columnwidth]{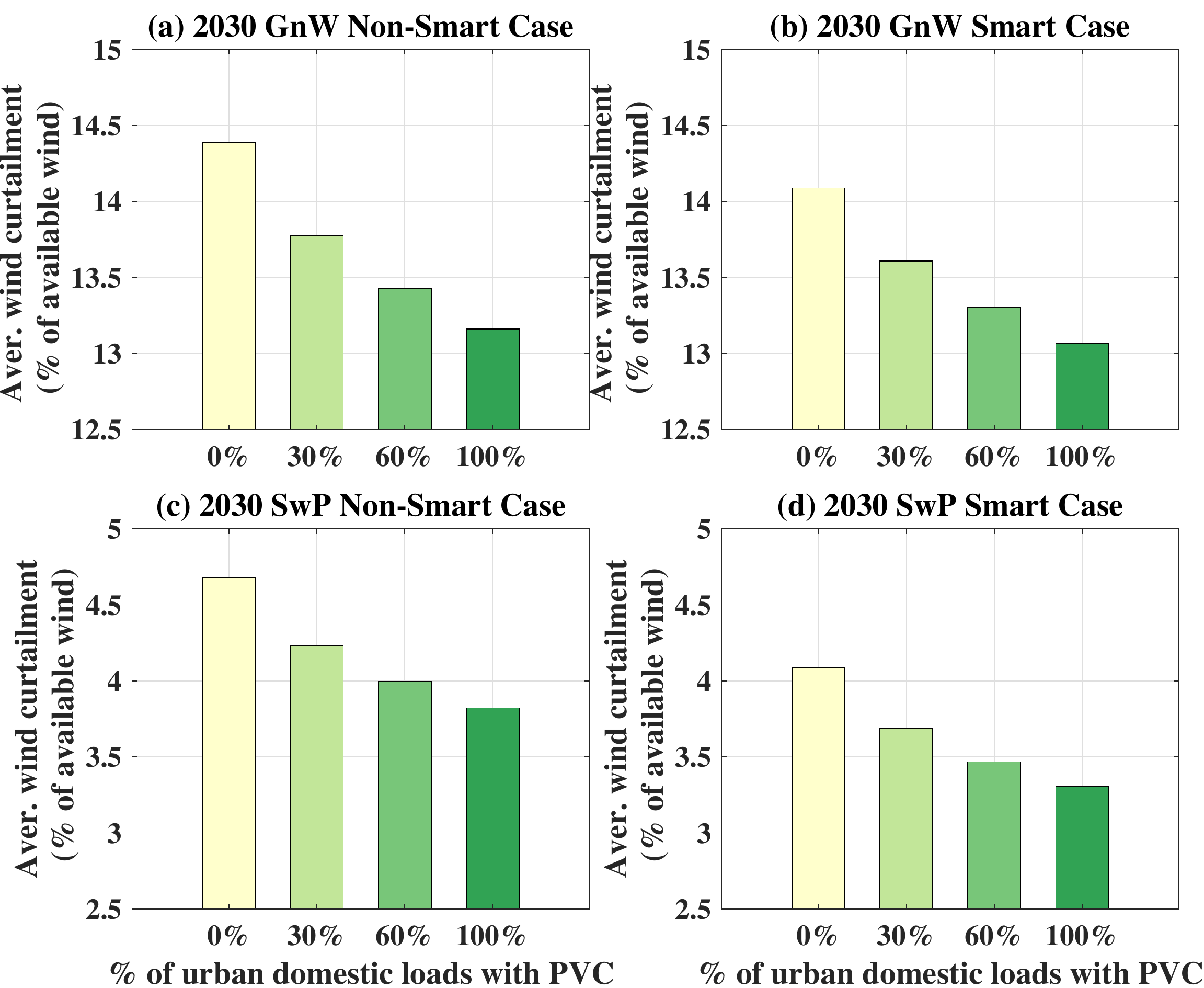}		
		\caption{Average wind curtailment (\% of available wind) for (a) 2030 GnW NSC; (b) 2030 GnW SC; (c) 2030 SwP NSC; (d) 2030 SwP SC}		
		\label{fig:WC}
	\end{figure}

 	\subsubsection{Wind Curtailment}
Besides the economic value, the benefits of EFR from PVC in wind accommodation are shown in Fig.~\ref{fig:WC}, in terms of average wind curtailment percentage with respect to the available wind power throughout the year. With 100\% of urban domestic loads equipped with PVC, the wind accommodation can be increased by about 1.1\% and 0.82\%, for the 2030 GnW and SwP scenarios. Although this number may seem marginal, it could increase the wind utilization up to around 3.3TWh per year for the 2030 GnW SC.

\begin{figure}[!t]
		\centering	
		\includegraphics[width=0.95\columnwidth]{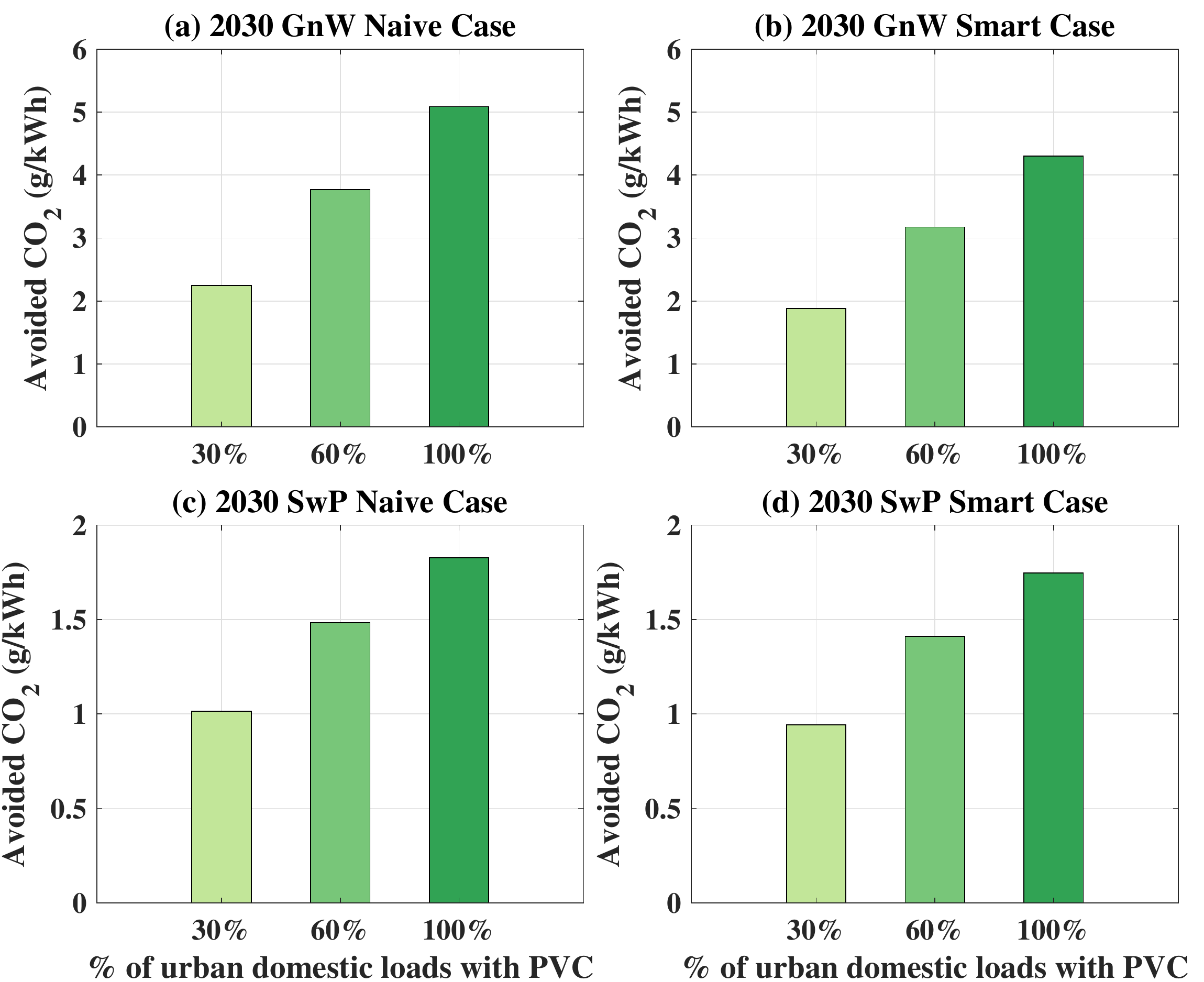}		
		\caption{Avoided $\rm CO_2$ emissions (g/kWh) for (a) 2030 GnW NSC; (b) 2030 GnW SC; (c) 2030 SwP NSC; (d) 2030 SwP SC}		
		\label{fig:CO2}
	\end{figure}

\subsubsection{$\rm CO_2$ Emission}
The environmental benefits of EFR from PVC are also evaluated regarding carbon emissions. The average system emission rate can be calculated by the ratio between total system emissions and the overall demand. The emission rates for each type of generation, as given in Table \ref{tab:char}, are considered to calculate this ratio, using the generation output for each thermal unit from the solution of the SUC. As in Fig.~\ref{fig:CO2}, the $\rm CO_2$ emission reduction for 2030 GnW can be 2 to 3 times than that in 2030 SwP. This can be explained by the fact that PVC is displacing traditional thermal generators for providing frequency response, and the requirement for response increases for higher wind penetration levels.

\begin{figure}[!t]
		\centering	
		\includegraphics[width=0.75 \columnwidth]{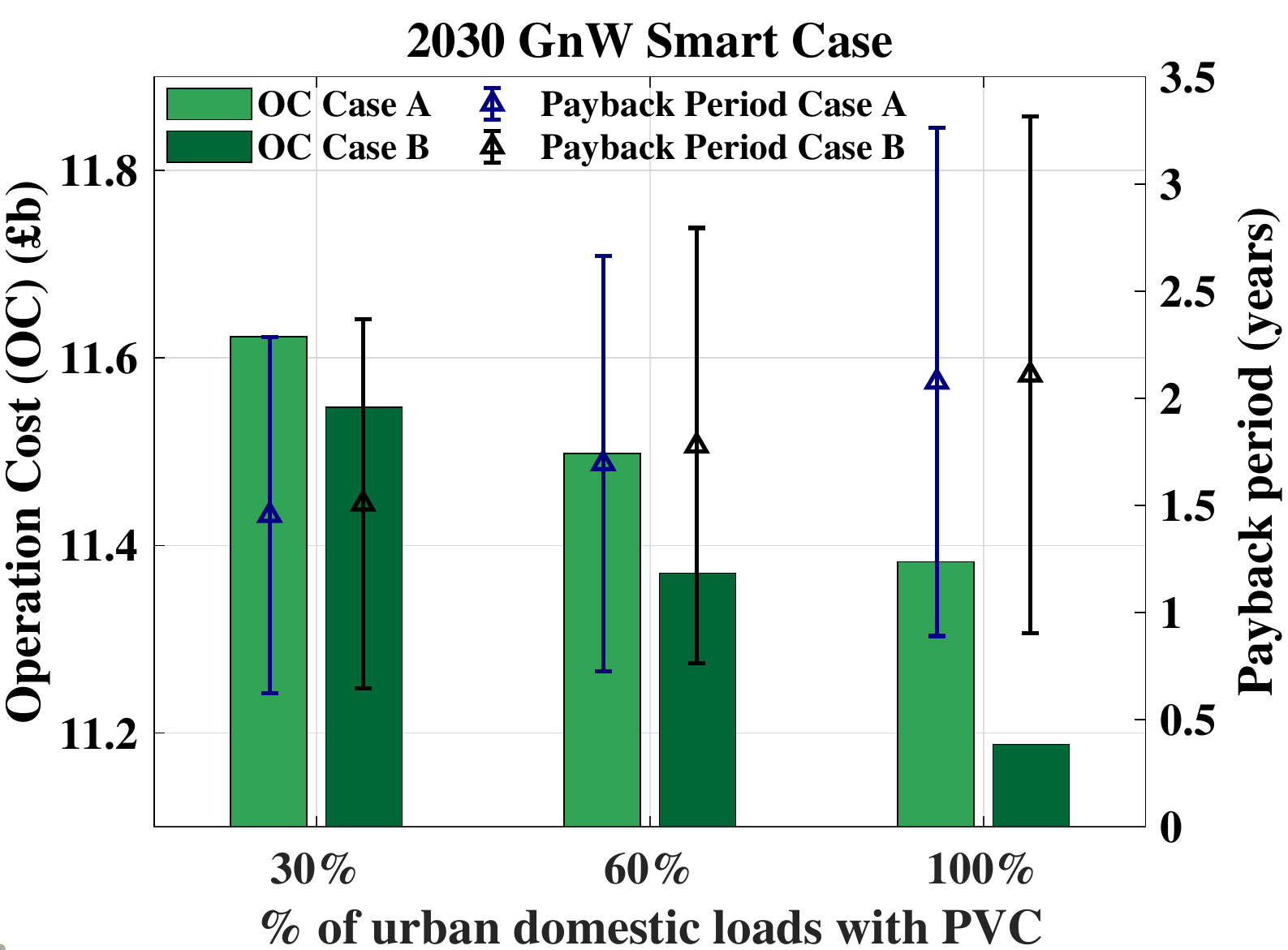}		
		\caption{\textcolor{m_blue}{Operation cost and payback period for 2030 GnW SC under normal (Case A) and fully controllable cases (Case B)}}		
		\label{fig:var}
	\end{figure}

\subsection{PVC in Fully Controllable Mode} \label{Sec:Controllable Case}

A comparison between provision of EFR from PVC under the normal mode (referred to as `Case A') and that under the fully controllable mode (referred to as `Case B') in terms of economic value for the 2030 GnW SC is presented in Fig.~\ref{fig:var}. 

The fully controllable mode allows to increase the power consumption beforehand in order to provide more EFR, at the expense of more energy consumption during a certain period of time. Intuitively, the increase in demand will give rise to energy cost as well. However, if the increased consumption can be supplied by extra wind generation with zero marginal cost , it can actually be cost-effective. This mode also enables demand with PVC to consume more flexibly, i.e. consume less when net-demand is high (less EFR required), which adds additional operation cost reduction compared with the normal mode. The above analysis is demonstrated in Fig.~\ref{fig:var}, where all Case B show a further reduction in operation cost with a maximum of \pounds 0.19b under 100\% PVC equipped level. It is interesting to note that the operation cost of Case B with 60\% PVC level is even less than that of Case A with 100\% PVC equipped level.

\begin{figure}[!t]
		\begin{center}		
		\includegraphics[width=0.95\columnwidth]{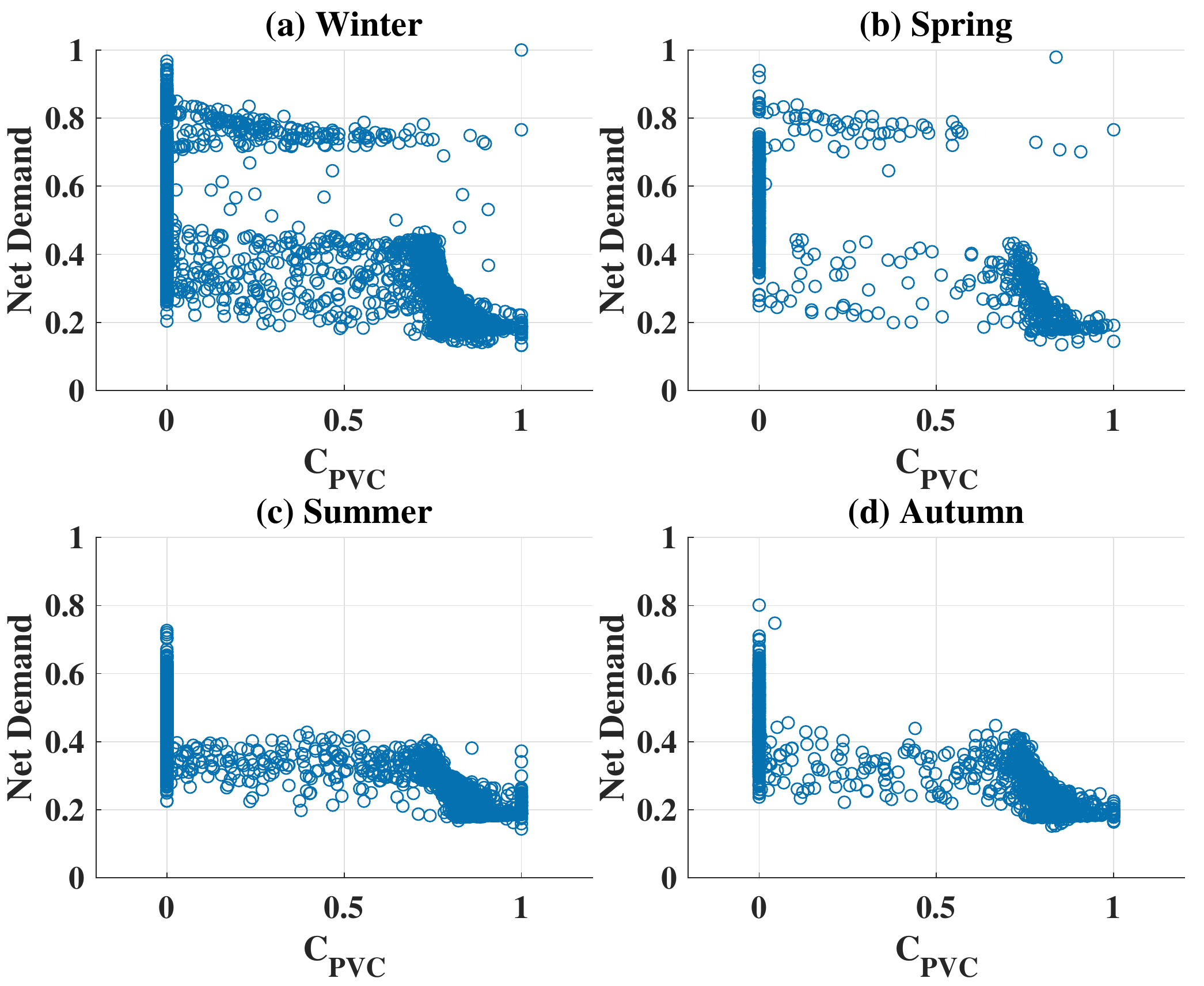}		
		\caption{Correlation between the urban domestic sector consumption and system net-demand (with 0.5GW BESS, 100\% of urban domestic loads with PVC) (a)winter; (b)spring; (c)summer; (d)autumn}		
		\label{fig:dv}
		\end{center}
	\end{figure}  

Although the fully controllable mode could bring in extra economic value, it also requires an additional investment cost, related to the increase of the required rating of PECs deployed at each CDC. The power processed by PECs when increasing the voltage to the maximum stipulated value can be calculated by replacing $V_{\rm min}$ in equation (\ref{eq:PEC}) with $V_{\rm max}$. And the final determined rating should take the maximum of these two cases, which leads to about 50\% more in rating requirement. This  results in a slightly longer payback period as in Fig.~\ref{fig:var}.

To further investigate this fully controllable mode, an index ($\rm C_{PVC}$) indicating the demand decision by the urban domestic sector is defined as in (\ref{eq:index}). $\rm P_{min}^{PVC}$ and $\rm P_{max}^{PVC}$ denote the minimum and maximum possible demand with PVC, which are calculated by the nominal demand, $\rm V_{min}$ and $\rm V_{max}$ along with the corresponding power-voltage sensitivities, neglecting the change in network and converter losses for simplicity. A value of $\rm C_{PVC}$ approaching 1 (0) implies the consumption decision approximating its maximum (minimum). 
\begin{equation} \label{eq:index}
 \mathrm{C}_\mathrm{PVC}=\frac{P_\mathrm{PVC}-\rm P^{PVC}_{min}}{\rm P^{PVC}_{max}-\rm P^{PVC}_{min}}  
\end{equation}

A correlation between $\rm C_{PVC}$ and system net-demand (normalized with respect to the maximum of the year) for each hour of the case with 100\% of urban domestic demand with PVC is presented in Fig.~\ref{fig:dv}. It can be seen that for a majority of cases, domestic customers with PVC are chosen to consume more when the net-demand is low, while consume less when the net-demand is relatively substantial. This is reasonable and cost-effective as the net-demand actually represents the demand required to be served by conventional thermal plants. When the net-demand is high, it is more likely that there are more thermal plants online to cover that net-demand, thereby requiring less EFR and vice versa.

However, there are some instances when the above philosophy is not perfectly followed. For example, the demand could choose a higher consumption even under relatively high net-demand (about 0.8) as the wind availability could drop significantly in the next hour, which leaves the thermal generators much less capable of providing frequency response. Fig.~\ref{fig:dv} shows these cases happen much more frequently during winter and spring when the wind energy is more volatile. 

	 Also, it is to be noted that $\textrm{C}_\textrm{PVC}=0$ is chosen for every value of net-demand within the vertical line shown in each sub-figure of Fig.~\ref{fig:dv} in which the net-demand of some instances can be very low. The reason behind is that in those cases, all the wind available has already been accommodated, which means increasing the consumption from PVC would also increase the cost of energy.


	\begin{figure}[!t]
		\begin{center}		
		\includegraphics[width=0.9\columnwidth]{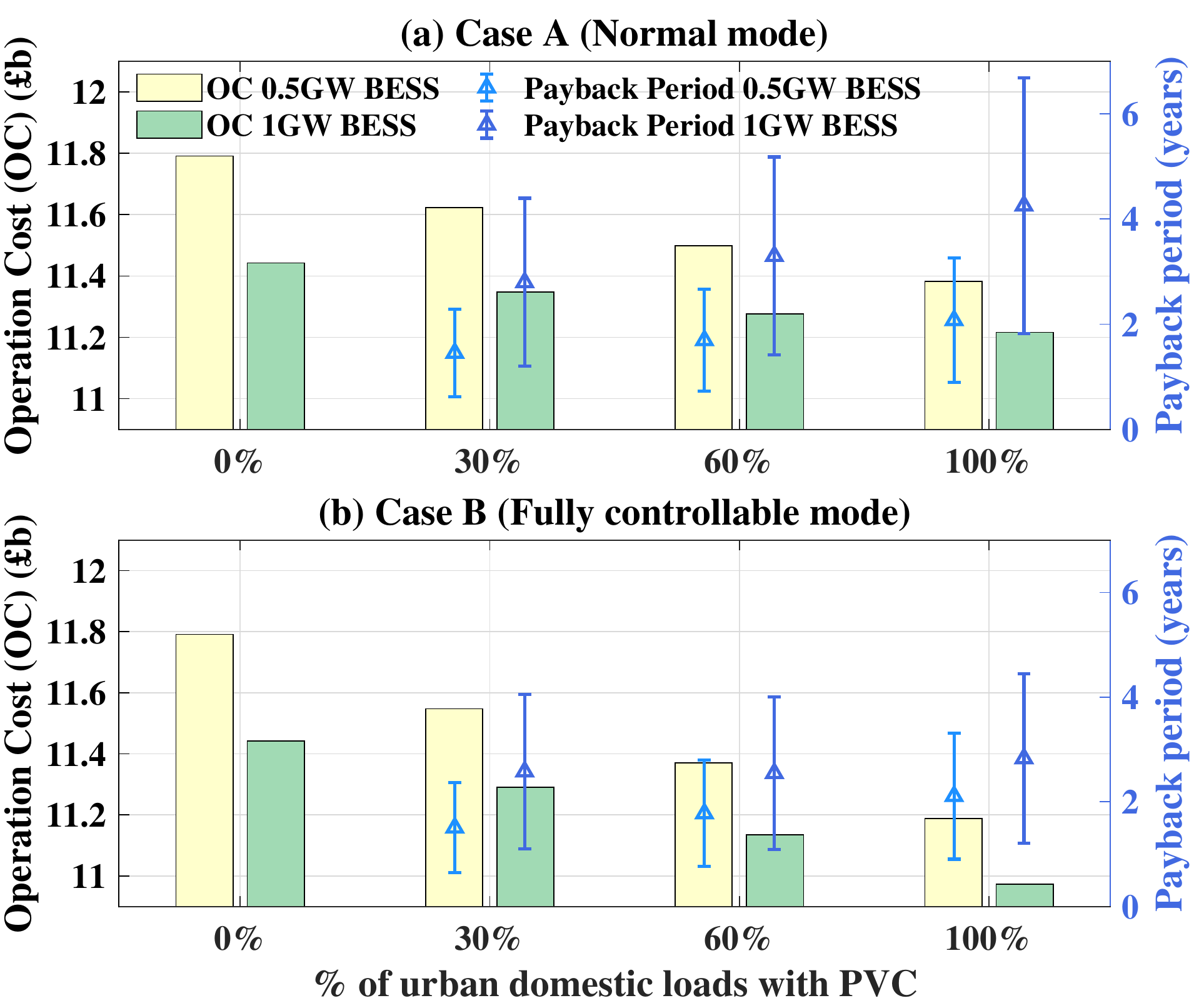}		
		\caption{\textcolor{m_blue}{Impact of BESS on the economic value of EFR provision from PVC, for two different ratings of the BESS: 0.5GW and 1GW}}		
		\label{fig:battery}
		\end{center}
	\end{figure}

\subsection{Impact of BESS} \label{Sec:storage impact}

BESS, as an competing technology, can also provide EFR. Currently, the GB power system holds around 0.48GW of battery storage providing this service \cite{NationalGridOutage}. A scenario with 1GW rating of BESS is hence presented in Fig.~\ref{fig:battery} to evaluate the impact of such competing resource. As expected, the operation cost savings from PVC utilization are reduced substantially (by about 45\% for all normal cases) when the BESS rating is doubled as in Fig.~\ref{fig:battery} (a). However, it is interesting to note that with 1GW BESS in the system, the fully controllable mode (`Case B') gains more economic value than that in case of 0.5GW BESS, which leads to a significant decrease in payback period as shown in Fig.~\ref{fig:battery} (b). This is due to the fact that the extra BESS provides more flexibility to the system, allowing the PVC to reduce the number of hours when it has to increase consumption for high net-demand conditions.

\section{Conclusions}

Point-of-load voltage control (PVC) in the urban domestic sector of Great Britain (GB) can provide up to 1.9 GW of enhanced frequency response (EFR) depending on the time of the day and the season. This results in annual savings in system operation cost of \pounds 0.23b and \pounds 0.72b for 2030 Slow Progression (SwP) and 2030 Green World (GnW) scenarios, respectively. The payback period for the investment in installing the power electronic compensators (PECs) for PVC varies \textcolor{m_blue}{between 0.3 to 6.7 years.} The payback period increases beyond 30\% penetration of PVC due to the diminishing value of EFR above a certain point. Increasing the demand for higher EFR provision or reducing it as necessary is shown to decrease the overall annual system operation cost by a further \pounds 0.19b for 2030 GnW scenario with only marginal increase in payback period. In this case, increasing the system-wide installed capacity of battery energy storage system (BESS) from 0.5GW to 1GW reduces the payback period further demonstrating the fact that PVC could effectively complement BESS towards EFR provision in the future GB power system. 

Further work is needed to quantify the EFR capability with PVC in presence of embedded generation (e.g. roof-top solar photovoltaic) and optimize the location and capacity of the PECs. Besides, the value assessment of PVC can be extended to include other services such as active network management and optimization of energy consumption.



\bibliographystyle{IEEEtran}

\bibliography{Value_Assessment}

\begin{IEEEbiography}[{\includegraphics[width=1in,height=1.25in,clip,keepaspectratio]{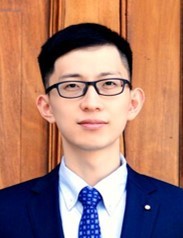}}]{Jinrui Guo (S'16)} received the B.Sc. and M.Sc. degrees from North China Electric Power University, Beijing, China, in 2012 and 2015, respectively. He is currently pursuing the Ph.D. degree from Imperial College London, U.K. His research interests include power system stability, smart grid, and renewable energy integration.\end{IEEEbiography}
 
\begin{IEEEbiography}[{\includegraphics[width=1in,height=1.25in,clip,keepaspectratio]{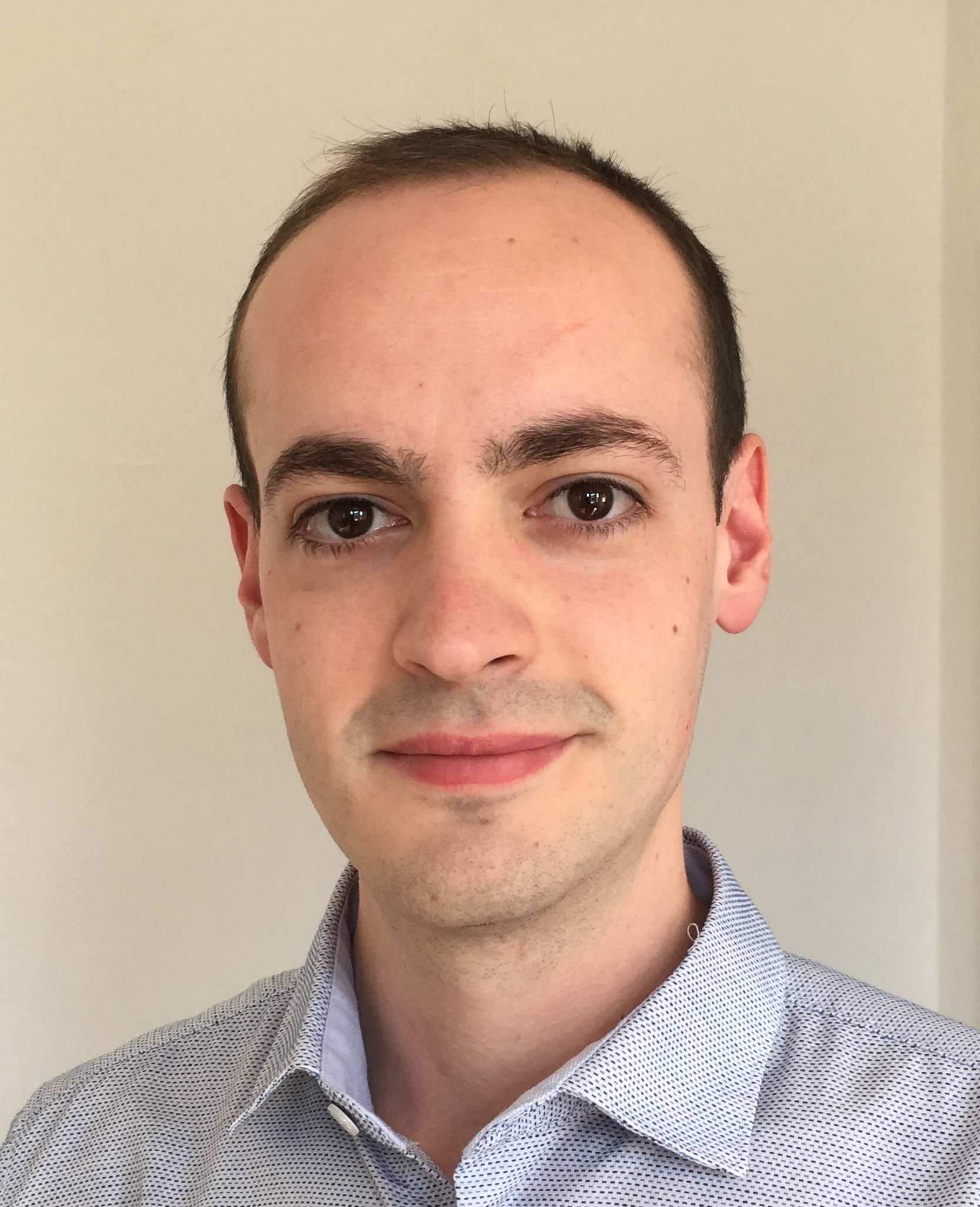}}]{Luis Badesa (S'14)} received the B.S. in Industrial Engineering degree from the University of Zaragoza, Spain, in 2014, and the M.S. in Electrical Engineering degree from the University of Maine, United States, in 2016. Currently he is pursuing a Ph.D. in Electrical Engineering at Imperial College London, U.K. His research interests lie in modelling and optimization for the operation of low-carbon power grids.\end{IEEEbiography}

\begin{IEEEbiography}[{\includegraphics[width=1in,height=1.25in,clip,keepaspectratio]{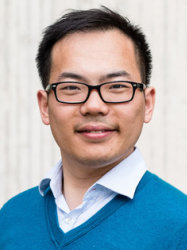}}]{Fei Teng (M'15)} received the BEng in Electrical Engineering from Beihang University, China, in 2009, and Ph.D. degree in Electrical Engineering from Imperial College London, U.K., in 2015. Currently, he is a Lecturer in the Department of Electrical and Electronic  Engineering, Imperial  College London, U.K. His research focuses on low-inertia power system operation, resilient cyber-physical energy systems, and objective-based data analytics for future energy systems.\end{IEEEbiography}

\begin{IEEEbiography}[{\includegraphics[width=1in,height=1.25in,clip,keepaspectratio]{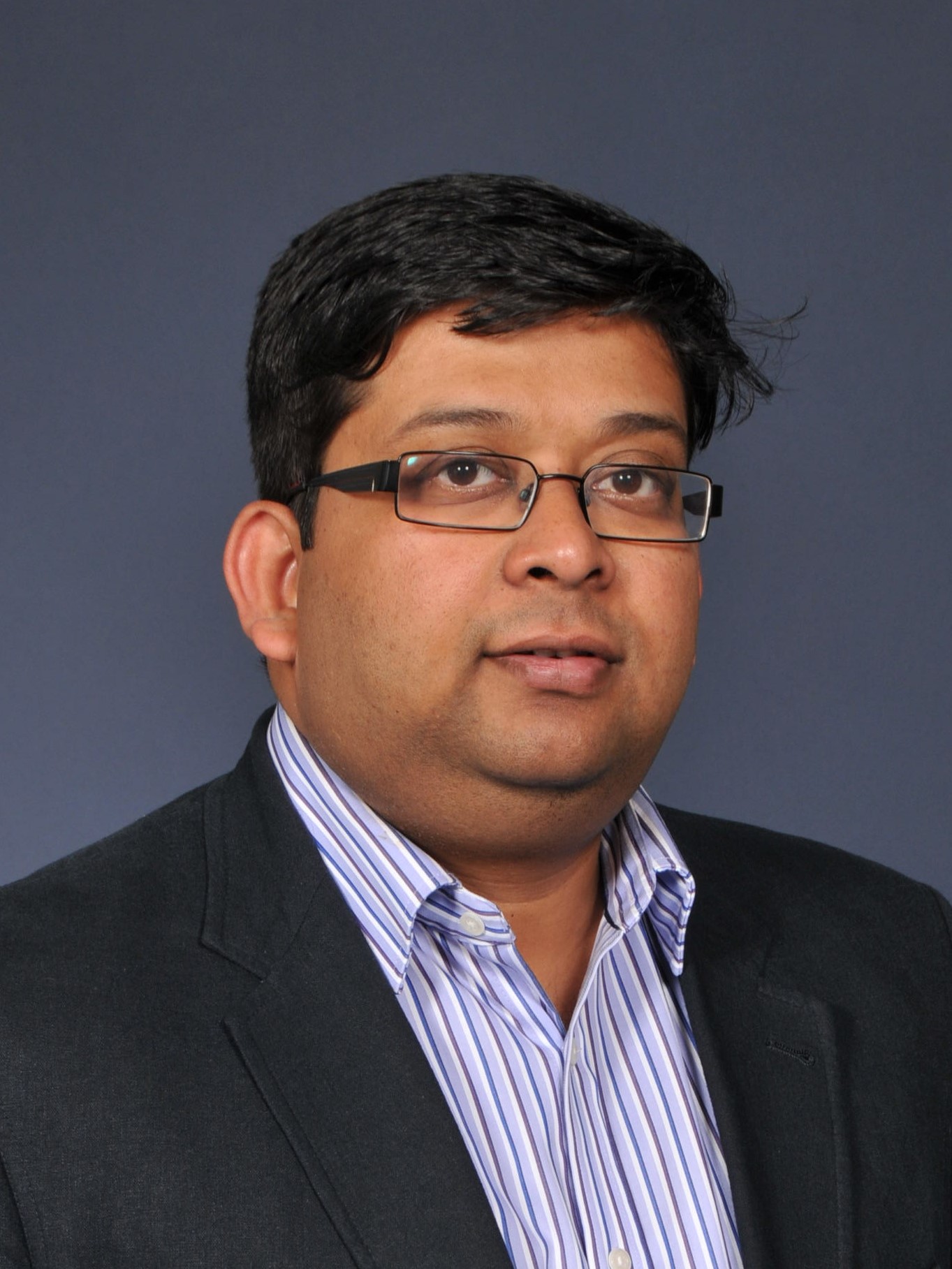}}]{Balarko Chaudhuri (M'06-SM'11)} received the Ph.D. degree in Electrical and Electronic engineering from Imperial College London, London, U.K., in 2005 where he is currently a Reader in Power Systems at the Control and Power Research Group. His research interests include power systems stability, grid integration of renewables, HVDC, FACTS, demand response and smart grids. Dr Chaudhuri is an editor of the IEEE Transactions on Smart Grid and an associate editor of the IEEE Systems Journal and Elsevier Control Engineering Practice. He is a Fellow of the Institution of Engineering and Technology (IET) and a member of the International Council on Large Electric Systems (CIGRE).
\end{IEEEbiography}
\begin{IEEEbiography}[{\includegraphics[width=1in,height=1.25in,clip,keepaspectratio]{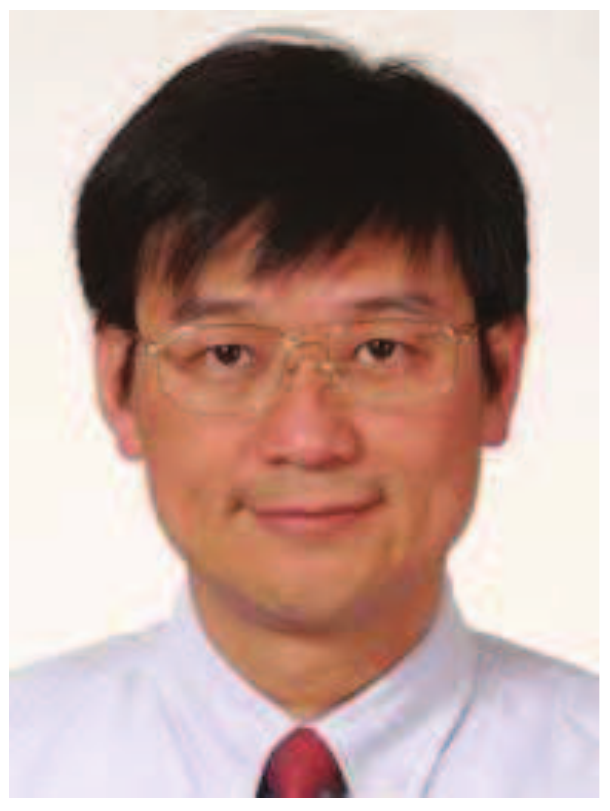}}]{Shu Yuen Ron Hui (M'87-SM'94-F'03)} received the B.Sc. (Hons.) degree in Electrical and Electronic engineering from the University of Birmingham, Birmingham, U.K., in 1984, and the D.I.C. and Ph.D. degrees from Imperial College London, London, U.K., in 1987.
He currently holds the Philip Wong Wilson Wong Chair Professorship with the University of Hong Kong, Hong Kong. Since 2010, he has concurrently held a part-time Chair Professorship in Power Electronics at Imperial College London. He has published over 200 technical papers, including about 170 refereed journal publications and book chapters. Over 55 of his patents have been adopted by industry.
Prof. Hui was the recipient of the IEEE Rudolf Chope Research and Development Award from the IEEE Industrial Electronics Society and the IET Achievement Medal (Crompton Medal) from the Institution of Engineering and Technology in 2010. He is a Fellow of the Australian Academy of Technological Sciences and Engineering. He is also the recipient of the 2015 IEEE William E. Newell Power Electronics Award. He is an Associate Editor of the IEEE TRANSACTIONS ON POWER ELECTRONICS and the IEEE TRANSACTIONS ON INDUSTRIALELECTRONICS.\end{IEEEbiography}

\begin{IEEEbiography}[{\includegraphics[width=1in,height=1.25in,clip,keepaspectratio]{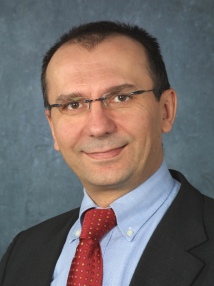}}]{Goran Strbac (M'95)} is a Professor of Electrical Energy Systems at Imperial College  London,  U.K.  His  current  research  is  focused  on  optimisation of energy systems’ operation and investment, energy infrastructure reliability and future energy markets.\end{IEEEbiography}

\end{document}